\documentclass[11pt]{article}
\usepackage{aaspp4,lscape,psfig}
\tighten

\def\gtsima{$\, \buildrel > \over \sim \,$}
\def\ltsima{$\, \buildrel < \over \sim \,$}
\def\simgt{\lower.5ex\hbox{\gtsima}}
\def\simlt{\lower.5ex\hbox{\ltsima}}

\begin{document}

\bigskip
\bigskip
\bigskip

\title{The MACHO Project 9 Million Star Color-Magnitude Diagram \\
 of the Large Magellanic Cloud}

\bigskip

\author{
      C.~Alcock\altaffilmark{1,16}, 
      R.A.~Allsman\altaffilmark{2},
      D.R.~Alves\altaffilmark{1,3,4},
      T.S.~Axelrod\altaffilmark{5},
      A.~Basu\altaffilmark{1},
      A.C.~Becker\altaffilmark{6},\\
      D.P.~Bennett\altaffilmark{4,7,16},
      K.H.~Cook\altaffilmark{1,4,16},
      A.J.~Drake\altaffilmark{5},
      K.C.~Freeman\altaffilmark{5},
      M.~Geha\altaffilmark{1},
      K.~Griest\altaffilmark{8,16},\\
      L.~King\altaffilmark{7},
      M.J.~Lehner\altaffilmark{9},
      S.L.~Marshall\altaffilmark{1,4},
      D.~Minniti\altaffilmark{1,10}, 
      C.~Nelson\altaffilmark{11},
      B.A.~Peterson\altaffilmark{5},\\
      P.~Popowski\altaffilmark{1},
      M.R.~Pratt\altaffilmark{12},
      P.J.~Quinn\altaffilmark{13},
      C.W.~Stubbs\altaffilmark{6,16},
      W.~Sutherland\altaffilmark{14,16},\\
      A.B.~Tomaney\altaffilmark{6},
      T.~Vandehei\altaffilmark{8},
      D.L.~Welch\altaffilmark{15}
}

\begin{center}
{\bf (The MACHO Collaboration)}
\end{center}

\bigskip

{\footnotesize
 
\altaffiltext{1}{Lawrence Livermore National Laboratory, Livermore, CA 94550}
 
\altaffiltext{2}{Supercomputing Facility, Australian National University,
        Canberra, ACT 0200, Australia }
 
\altaffiltext{3}{Space Telescope Science Institute, 3700 San Martin Dr.,
Baltimore, MD, 21218}
 
\altaffiltext{4}{Visiting astronomer at the Cerro Tololo Intra-American
Observatory}
 
\altaffiltext{5}{Mt.~Stromlo and Siding Spring Observatories, ANU,
        Weston Creek, ACT 2611, Australia}
 
\altaffiltext{6}{Departments of Astronomy \& Physics,
        University of Washington, Seattle, WA 98195 }
 
\altaffiltext{7}{Physics Department, University of Notre Dame, Notre
        Dame, IN 46556 }
 
\altaffiltext{8}{Department of Physics, University of California,
        San Diego, La Jolla, CA 92093 }
 
\altaffiltext{9}{Department of Physics, University of Sheffield,
        Sheffield S3 7RH, UK }
 
\altaffiltext{10}{Departmento de Astronomia,
P. Universidad Catolica, Casilla 104, Santiago 22, Chile}
 
\altaffiltext{11}{Department of Physics, University of California,
        Berkeley, CA 94720 }
 
\altaffiltext{12}{Center for Space Research, MIT, Cambridge MA 02139 }
 
\altaffiltext{13}{European Southern Observatory, Karl-Schwarzchild Str. 2,
        D-85748, Garching, Germany }
 
\altaffiltext{14}{Department of Physics, University of Oxford,
        Oxford OX1 3RH, U.K. }
 
\altaffiltext{15}{Departments of Physics \& Astronomy,
   McMaster University, Hamilton, Ontario, Canada L8S 4M1. }
 
\altaffiltext{16}{Center for Particle Astrophysics,
        University of California, Berkeley, CA 94720}
}

\clearpage
\begin{abstract}

We present a 9 million star color-magnitude diagram (9M~CMD)
of the Large Magellanic Cloud (LMC) bar.  The 9M~CMD reveals a complex 
superposition of different age and metallicity stellar populations, with
important stellar evolutionary phases occurring over three 
orders of magnitude in number density.
First, we count the non-variable red and blue supergiants, the
associated Cepheid variables, and measure the stellar
effective temperatures defining the Cepheid instability strip.
Lifetime predictions of stellar evolution theory are
tested, with implications for the origin of low-luminosity Cepheids.
The highly-evolved
asymptotic giant branch (AGB) stars in the 9M~CMD
have a bimodal distribution in brightness,
which we interpret as discrete old populations ($\simgt$1~Gyr).
The faint AGB sequence may be metal-poor and very old.
Comparing the mean properties of giant branch
and horizontal branch (HB) 
stars in the 9M~CMD to those of clusters,
we identify NGC~411 and M3 as templates
for the admixture of old stellar populations in the bar. 
However, there are several indications that the old and metal-poor
field population has a red HB morphology:
the RR~Lyrae variables lie preferentially on the
red edge of the instability strip, the AGB-bump is very red, 
and the ratio of AGB-bump stars to RR~Lyraes is quite large.
If the HB second parameter is age,
the old and metal-poor field population
in the bar likely formed after the oldest LMC clusters.
Lifetime predictions of stellar evolution theory
lead us to associate a significant 
fraction of the $\sim$1 million red HB clump giants
in the 9M~CMD with the same old and metal-poor population
producing the RR~Lyraes and the AGB-bump.
In this case, compared to
the age-dependent luminosity predictions of stellar evolution theory, 
the red HB clump is too bright relative to the RR~Lyraes and AGB-bump.
Last, we show that the surface
density profile of RR~Lyraes is fit by an exponential,
favoring a disk-like rather than spheroidal distribution.
We conclude that the age of the LMC disk is probably similar to the
age of the Galactic disk.

\end{abstract}

\keywords{
galaxies: Magellanic Clouds, stellar content --- 
stars: color-magnitude diagrams (HR diagram), 
evolution, horizontal-branch, supergiants --- 
stars: variables: AGB, Cepheids, RR Lyrae variable 
}

\clearpage

\section{ Introduction }

The Large Magellanic Cloud (LMC) is a nearby galaxy
with one spiral arm and a bar (de~Vaucouleurs 1954).
The LMC exhibits a disk-like exponential
surface brightness profile.  The surface brightness profile
of the LMC bar, unlike typical galactic bars,
is also exponential (Bothun \& Thompson 1988).
The study of the resolved stellar populations in the
LMC bar is challenging because the 
surface density of stars is quite high.
However, the proximity of the LMC otherwise favors 
detailed observations.
It is important to 
study the stellar populations 
of nearby galaxies like the LMC
in order to understand the processes of galaxy evolution.
In particular, the formation of exponential disks 
is an outstanding problem in cosmogony
(e.g.~Freeman 1970, Fall \& Efstathiou 1980, Dalcanton et al.~1997). 

Butcher's (1977) seminal study of the main sequence luminosity function 
in the LMC
showed that the star formation histories of the LMC disk
and the solar neighborhood Galactic disk are different.
Butcher (1977) concluded that
the bulk of LMC field stars formed $\sim$3-5~Gyr ago 
rather than $\sim$10~Gyr ago,
when the Galactic disk appeared to have formed (Salpeter 1955).
Numerous subsequent studies of LMC field stars
have supported this
initial claim (Bertelli et al.~1992, Westerlund 1997, van den Bergh 1998).  
For example, Geha et al.~(1998) found that roughly half of the stars 
in outer disk fields of the LMC formed before and after 4 Gyrs ago.
Recent stellar populations studies,
such as those made with the {\it Hubble Space Telescope} 
(e.g.~Geha et al.~1998, Olsen~1999, Holtzman et al.~1999), 
are now beginning to probe the spatial
variations of the LMC star formation history in detail.
Holtzman et al.~(1999) argue that the LMC bar
has a larger relative component of older stars than the outer disk fields.

The first color-magnitude diagram (CMD)
study of the LMC bar was made by Tifft and Snell (1971).
Their CMD contained $\sim$1000 stars to a limiting
magnitude of $V \approx 18$ mag.  Tifft and Snell (1971) first observed the
tip of the red giant branch (RGB) in the LMC bar,
thus identifying an old stellar population.
Hardy et al.~(1984) obtained a CMD of the LMC bar with $\sim$18000 stars
to a limiting magnitude of $V \approx 21$ mag.  They argued that the
dominant stellar population in the bar formed 
$\sim$1-3 Gyr ago, and concluded that an older population in their CMD
was only a weak component.  Elson et al.~(1998)
reached a similar conclusion.
They analyzed a {\it Hubble Space Telescope\,} CMD of an ``inner disk'' field
(close to the bar) and argued that the bar
formed $\sim$1 Gyr ago and the disk formed $\sim$3 Gyr ago.
Olsen (1999) and Holtzman et al.~(1999) have contested the
relatively young disk and bar advocated by Elson et al.~(1997).

In this paper, we present a 9 million star 
color-magnitude diagram (9M~CMD) of the LMC bar.  Our fields cover almost
the entire LMC bar (10 square degrees) to a limiting
magnitude of $V \approx 22$ mag.  These are the same fields
extensively analysed for microlensing by Alcock et al.~(1997).
The 9M~CMD is assembled from
MACHO Project two-color instrumental photometry 
calibrated to the standard Kron-Cousins $V$ and $R$ system
(Alcock et al.~1999).
For these 9 million stars, the precision of the 
photometric calibration
is $\sigma_V$ = $\sigma_R$ = 0.02 mag.  
In addition to the sheer number of stars, near total spatial
coverage of the bar, and high precision calibration, 
each star in the 9M~CMD is represented by
a 6-year lightcurve consisting
of $\sim$1000 two-color photometric measurements.
Variable stars are easily identified with
these time-sampled data.
The 9M~CMD is the product of wide-field imaging array
detector technology and a dedicated, ground-based, 1-m class survey telescope.

In order to 
discuss the taxonomy of the 9M~CMD,
we begin with a ``tour.'' 
The tour is intended to give the reader an overall impression
and introduce key features which are the
subject of more detailed discussions that follow. 
After the tour, we turn our attention
to the young LMC stellar populations.
The core helium-burning red and blue 
supergiants and Cepheid variable stars
are particularly interesting examples of the late
stages of stellar evolution of intermediate-mass stars.
Supergiants in the LMC have a long history as testing grounds
for the theory of stellar evolution, although past attempts to study
these rare stars have sometimes proven difficult 
for the small samples available
(e.g.~Maeder \& Meynet 1989, Langer \& Maeder 1995).   
The 9M~CMD represents the largest homogeneous sample of non-variable
supergiants and Cepheids ever assembled, 
thus allowing for new and precise comparisons with theory.  

Next, we examine the old LMC stellar populations.
The metallicities and ages of these 
stars are not well-known, which is
an obstacle to the sort of detailed comparisons with theory we
make with the young stellar populations.  Therefore, 
a model for the old LMC stellar populations is first constructed.  
In order to minimize
the dependence of this analysis on purely theoretical results,
we rely primarily
on the comparison of major features in the 9M~CMD 
and the properties of variable stars with their counterparts in clusters.
Clusters are useful because they serve as ``template''
populations, or building blocks for the composite 9M~CMD.
Our simple model for the old LMC populations is intended to 
serve as a check of more sophisticated,
but isochrone-dependent, analyses of the LMC star formation history.
In many respects, 
we repeat our analysis of the intermediate-mass supergiants 
and Cepheids, but for the low-mass helium-burning giants.  
By studying the low-mass RR~Lyrae variables, we make inferences 
to the nature of the old and metal-poor LMC field population.
This elusive LMC population probes the formation epoch of the LMC, 
with general implications for cosmogony.

\section{Tour of the 9 Million Star Color-Magnitude Diagram}

\subsection{ Construction }

The MACHO photometry data
and transformation of these data to the 
Kron-Cousins $V$ and $R$ standard system
are discussed extensively by Alcock et al.~(1999), 
to which the reader is referred for further details.  
These calibrated photometry data may be properly compared to
other data on the Kron-Cousins standard system.  Moreover,
we may infer accurate effective temperatures and
bolometric luminosities with consideration
of stellar atmospheres convolved with the standard
$V$ and $R$ passbands,
thus allowing for direct comparisons of stars in the 9M~CMD 
with theoretical results.  
Unless otherwise noted, the photometry data analysed here are derived 
from single observations; they are not time-averaged magnitudes
and colors.

The 9 million stars
analysed in this paper are distributed throughout 22 
MACHO Project survey fields.
A map of these fields
overlayed on a wide-field image of the LMC
(Bothun and Thompson 1988) is presented
in Alcock et al.~(1997; their Figure~1).   Coordinates of the
field centers are also provided. 
For each of the 22 fields, 
we bin the photometry data into a Hess diagram.  A Hess
diagram is a CMD that also contains information on the number of stars
as a function of color and magnitude.
We use a bin size of 0.01 mag in $(V-R)$ and
0.02 mag in $V$ with axes running from $-0.5 < (V-R) < 1.5$ and
$22 < V < 12$ mag.  
We make the final 9M~CMD ``image'' by summing the
22 individual field CMDs using standard 
IRAF\footnote{The Image Reduction and Analysis
Facility, v2.10.2, operated by the National Optical
Astronomy Observatories.} image processing routines.

\subsection{ The Tour }

We present the 9M~CMD in Figure~1.  
The image is log-scaled and color coded.
The log of the number of stars in each pixel 
increases following the
sequence: $blue \rightarrow green \rightarrow yellow \rightarrow red$.
Important stellar evolutionary phases occur over
three orders of magnitude in stellar number density.
The highest pixel value in this image is $\sim$3.5
dex, which is found at the peak of the red horizontal branch clump 
(feature C in Fig.~1).
This single high pixel therefore represents $\sim$3000 stars
with the same $(V-R)$ color and $V$ mag to within the 
resolution of our adopted pixel/bin size.  
Figure~1 is designed to give an overall 
impression of the 9M~CMD data.  A
small number of stars brighter than $V = 12$ mag are not shown; this is
above the saturation limit of the MACHO image data for the LMC anyway.
Incompleteness is clearly evident for $V > 21$ mag.
For example, the number density of main sequence sequence stars 
would still be increasing at this brightness 
if not for incompleteness in the data
(Alcock et al.~1999).  We will restrict
our analyses to the brighter stars in the 9M~CMD.  
We have labeled nine features with the letters
{\bf (A)} through {\bf (I)} in Figure~1.  
These are identified as follows.

{\bf (A) The main sequence.} 
These stars are primarily
core hydrogen-burners
(Maeder \& Meynet~1989).
The majority of main sequence stars visible in the 9M~CMD have 
either O, B, or A spectral types.
The ridge line of the main sequence is not exactly vertical,
but instead runs slightly from blue to red with decreasing brightness.
There are almost no stars blueward of the upper main sequence.
The stars bluer than the main sequence ridge line 
and increasing in number at progressively
fainter magnitudes are consistent with our photometric errors 
and the total numbers of stars found on the main sequence.
The range of brightnesses on the main sequence may be interpreted
as a range of initial masses.  The bright upper main
sequence stars indicate recent star formation in the LMC bar;
these stars only live $\sim$20 Myr (Maeder \& Meynet~1989).

{\bf (B) The giant branch.} 
The stars on the giant branch are old, but in a mix of different
evolutionary phases.
Most of these stars are on the first-ascent red giant branch (RGB);
they have degenerate helium cores
and hydrogen-burning shells (Schwarzschild 1958).
The base of the RGB (i.e. the subgiant branch) is not visible
in the 9M~CMD (it is too faint). 
However, the termination, or ``tip''
of the RGB is seen at feature {\bf(E)}.
Stars at the tip of the RGB ignite helium in their cores and
evolve very rapidly to the horizontal branch, near feature {\bf(C)}.
Some of the stars on the giant branch are also on the 
asymptotic giant branch (AGB).  These stars are helium shell-burners,
and are in an evolutionary state more advanced than the horizontal branch.
For many stars in the 9M~CMD, the
AGB begins at {\bf(D)}, 
continues past {\bf(E)} and into region {\bf(F)}.
We do not resolve the RGB and AGB in the 9M~CMD.
Stars located redward of the giant branch
ridge line are affected by differential 
reddening or photometric
errors.  In addition, some of these stars are foreground
stars (in our Galaxy), and some are galaxies behind the LMC.

The association of the LMC bar giant branch with an ``old'' population
was first made by Tifft and Snell (1971).  This is 
more precisely defined as a population older than $\sim$1 Gyr,
a characteristic age also known as 
the RGB phase transition (Sweigart et al.~1990).  
The initial mass of stars at this transition is likely
$\sim$2 $M_{\odot}$ (Bertelli et al.~1985, Sweigart et al.~1990),
which sets a lower limit on the age of a stellar
population with a fully developed (extended) giant branch.

{\bf (C) The horizontal branch red clump.} 
The stars in the horizontal branch (HB) red clump are primarily 
core helium-burners older
than $\sim$1 Gyr (Seidel et al.~1987), although some of these
stars may be in different evolutionary phases. 
The red HB clump is very prominent in the 9M~CMD.  
Note the elongation along the reddening vector due to
differential reddening in the LMC (like some of the stars redward of the
giant branch ridge line).   There are also 
brighter and bluer red HB clump stars, 
visible as a faint extension of the
red HB clump running toward label {\bf(A)}.
These are unresolved blends\footnote{The
chance superposition of two stars 
on the sky which we are unable to resolve with our 
ground-based image data.  Some of these may be binary
systems, also unresolved in our data.} of red clump and main sequence stars.
Although confused with the giant branch, clump-clump 
and clump-giant branch blends are likely present 
in similar numbers  in the 9M~CMD.

The ages and metallicities of the
red HB clump giants in the LMC are surprisingly
ill-constrained.
Hardy et al.~(1984) argued that
the red HB clump giants in the LMC bar are most likely
$\sim$1 to 3 Gyrs old.  However, their upper age limit
is based on uncertain subgiant branch starcounts.
In this work, we will constrain the ages and metallicities
of the red HB clump giants 
using other stellar evolutionary features in the 9M~CMD,
such as the giant branch {\bf(B)},
the AGB-bump {\bf(D)}, and the properties
of the field RR Lyrae variable stars.

It is useful to compare the red HB clump giants to the RR~Lyrae
variable stars in the 9M~CMD because they are closely related
in a stellar evolutionary sense, and the characteristic ages
of RR~Lyraes are known. 
RR~Lyraes are believed to be very old, having been
found only in very old clusters ($\simgt$9 Gyr).  In the LMC, RR~Lyrae
are also known to be
metal-poor (see Olszewski et al.~1996 for a review).  We find the red
HB clump is $\Delta V = -0.28$ mag brighter than the mean
brightness of the RR Lyraes
in the 9M~CMD (the RR Lyraes are not distinct in Figure~1,
but see \S4.2 of this paper).  
This brightness difference is
larger than expected on theoretical grounds for an old, coeval HB
(e.g. Fusi Pecci et al.~1996).
However, this alone is not proof that the RR Lyraes
and red HB clump giants in the LMC bar have different ages.
The luminosity predictions of stellar evolution theory are 
very difficult to test at this
high level of precision (Alves \& Sarajedini 1999).
If we estimate the mean metallicity of the 
red HB clump giants in the LMC bar, and assume that $age$
is the so-called second parameter influencing HB morphology
(Sandage \& Wiley 1966, Stetson et al.~1996),
then the color, brightness,
and number of red HB clump giants may give an indication to their age
(Lee et al.~1994, Sarajedini et al.~1995, Hatzidimtriou 1991).

The analog of red HB clump stars for populations
younger than the RGB phase transition 
are also present here.
These stars populate a feature in the 9M~CMD that has 
been called the VRC, for vertically-extended 
red clump (Zaritsky \& Lin 1997).  The VRC has been observed in 
other local group galaxies, such as Carina (Smecker-Hane et al.~1994)
and Fornax (Saviane et al.~1999).
If we define the VRC as a CMD feature populated by stars in the same
stellar evolutionary phase, then
the VRC in the LMC is composed of $\sim$2 to 4 $M_{\odot}$ core helium-burning
red giants (Corsi et al.~1994).
The VRC is the high-mass extension 
of the red HB clump and the 
low-mass extension of the red supergiants,
feature {\bf (I)}.
The VRC is visible in the 9M~CMD just
to the left (blueward) of the label {\bf(B)} in Figure~1.  
At this brightness,
the giant branch is beginning to ``turn over'' to redder colors
which makes it easier to distinguish the VRC sequence.

VRC stars with initial masses very close to the RGB phase transition
may actually be up to $\sim$0.5 mag fainter 
than the peak concentration of the red HB clump in the 9M~CMD
(Corsi et al.~1994, Girardi 1998). Bica et al.~(1998) and
Piatti et al.~(1999) have probably
observed this ``sub-clump'' in several outer disk fields of the LMC.
After reaching a minimum brightness,
progressively younger stars in this evolutionary phase will then
be brighter.  

The density of stars along the VRC sequence will depend 
on the lifetimes of red giants of different initial masses,
and also on 
the recent star formation history of the LMC.  
The theoretical lifetimes of VRC stars are very
sensitive to the parameterization of convective
overshoot, which may depend on initial mass and metallicity
in a non-trivial manner (Schroder et al.~1997).
Stochastic star formation throughout the LMC disk
may also cause variations in the density of stars
along the VRC, complicating comparisons with theory.
Bica et al.~(1998; but see Piatti et al.~1999) 
and Zaritsky and Lin (1997) have
attributed stellar density variations along the VRC 
to extragalactic stellar populations.
The VRC as defined here is a well-known 
stellar evolutionary branch
in CMDs representing mixed-age populations, and should not
be confused with the fluctuations in stellar 
density along this branch that have lead to claims of
new stellar populations in front (or behind) the LMC.

{\bf (D) The asymptotic giant branch bump.} 
We find a small concentration of stars on the giant branch 
approximately one mag brighter than the horizontal branch.
This feature was first observed
by Hardy et al.~(1984)
in their CMD of the LMC bar.
It is the AGB-bump, a 
slight evolutionary pause marking the transition from 
core to shell helium-burning 
(Castellani, Chieffi \& Pulone 1991).
The AGB-bump is the ``base'' of the AGB.  
The AGB-bump stars in the 9M~CMD 
are likely the same old population as the RR Lyrae stars, otherwise
this feature would be much brighter (Alves \& Sarajedini 1999).

{\bf (E) The tip of the red giant branch.} 
The tip of the red giant branch is defined by ignition 
of the degenerate helium core in old (low-mass) stars,
an event known as the helium flash (Renzini \& Fusi Pecci 1993).   
In the 9M~CMD, some stars at the blue edge of the
tip of the RGB are likely unresolved blends
of two giant branch stars.  These blends
``puff up'' the tip of the RGB.

{\bf (F) The asymptotic giant branch.} 
These bright and very red stars are on the AGB.
The AGB consists of two
major helium shell-burning evolutionary phases: 
the ``early AGB'', during which the outer hydrogen
shell is extinguished, and the ``thermal-pulsing AGB,''
marked by the reignition of the hydrogen shell
(Iben \& Renzini 1983).
The transition from the early AGB to the thermal-pulsing AGB 
is theoretically predicted
to occur near the tip of the RGB, and
may be associated with the onset of pulsation in these stars
(Alves et al.~1998, Wood et al.~1998).   
The AGB stars brighter or redder than the tip of the RGB are all likely 
experiencing thermal pulses.  The label {\bf(F)}
in Figure~1 marks the upper envelope of where most of the
AGB stars are found.  The AGB stars populate
the region between the label \& arrow {\bf(E)} and up to the label {\bf(F)}.  
At the red edge of the AGB region (where the most highly-evolved
AGB stars reside), the 9M~CMD reveals a bimodal
structure, which we will examine in detail.

{\bf (G) The blue supergiants.} 
These are $4 \simlt M \simlt$ 9 $M_{\odot}$ core helium-burning
stars.  They spend most of their
post-main sequence lifetimes as either blue or red supergiants
(Maeder \& Meynet 1989).
While ``looping'' between
the red and blue phases, these supergiants cross the instability strip
and become Cepheid variable stars.
The blue supergiants therefore define the ``tips'' of the
blue loops.  This is the lowest density stellar evolution feature 
we identify in the 9M~CMD.
The sequence runs from red to blue with increasing brightness and is visible
between the main sequence and feature {\bf(H)}, the foreground
Galactic disk stars.

{\bf (H) The foreground Galactic disk stars.} 
The wide-field coverage of the 9M~CMD results in significant
``contamination'' by Galactic foreground stars.
The foreground Galactic disk stars 
lie along sequence {\bf(H)} in the 9M~CMD.
The Galactic spheroid and thick disk make a small contribution
at this brightness in the 9M~CMD (Yoshii \& Rodgers~1989).
The moderately low Galactic latitude of the LMC
($b \approx -30^{\circ}$) increases the number of foreground disk
stars relative to halo and thick disk stars over the case of starcounts
at the Galactic poles.

{\bf (I) The red supergiants.} 
These are the $4 \simlt M \simlt$ 9 $M_{\odot}$
core helium-burning red giant stars 
associated with the blue supergiants 
and the Cepheid variables.
They are also the high-mass extension of the VRC.
Inspection of Figure~1 shows that there are more 
red than blue supergiants in the LMC bar.  
We will quantify this observation 
in the next section.

\section{ Young Stellar Populations: Supergiants and Cepheids }

The supergiant and VRC sequences in the 9M~CMD
are collectively known as ``intermediate-mass'' stars.  
Intermediate-mass stars
ignite helium non-degenerately, 
but develop a highly electron-degenerate
carbon-oxygen core after exhaustion of helium. 
The exact mass range depends on chemical composition, but is
likely $\sim$2 to 9 $M_{\odot}$ (Bertelli et al.~1985).  
After these stars leave the main sequence 
and evolve rapidly to become red giants (or supergiants, depending
on the initial mass of the star),
helium ignites in the core and a
hydrogen shell burns outward.   
This shell eventually approaches a chemical composition
discontinuity left by the furthest extent of the convective envelope
during core hydrogen-burning, which triggers
rapid movement of the star back to the blue side of the CMD
(Lauterborn et al.~1971, Stothers \& Chin 1991,
but see also Renzini et al.~1992).
Subsequently, these stars evolve back to the red side of the CMD, 
which marks the end of the first ``blue loop.''
The helium shell ignites (double shell-burning), and
these stars may make a second blue loop (Becker 1981).  Second
blue loops are even less well understood than first blue loops
(Hoeppner et al.~1978), but are theoretically predicted to last
$\sim$10\% of the lifetime of the first blue loop, if they occur at all.

The first blue loop is sometimes called the Cepheid loop, 
because it is the longer of the two loops and most
Cepheids are believed to be in this phase of evolution (Becker 1981).
Even during the first blue loop, these stars evolve very 
quickly between the tip of the blue loop and the red supergiant
sequence.  This rapid evolution manifests
as a ``gap'' in CMDs representing young stellar populations,
also known as the
Hertzsprung gap.  
Blue loops tend to
become shorter in lower mass stars until no loops occur
(Maeder \& Meynet 1989).  This 
stellar evolutionary trend is clearly confirmed by the
blue supergiant sequence in the 9M~CMD, which becomes redder
at lower luminosities.  We observe the
red supergiants making blue loops, but not the VRC stars.
The blue supergiant sequence in the 9M~CMD represents
the outer-envelope of intermediate-mass stars on blue loops.

While stellar evolution theory for intermediate-mass 
stars successfully 
predicts the major observational characteristics of supergiants
and Cepheids, the models may fail to match observation in detail
(Renzini et al.~1992; Langer \& Maeder 1995).
Therefore, it is worth
examining the supergiants in the 9M~CMD and making a comparison
with theory for the possibility of advancing the theory.
We assume that the intermediate-mass stars in the LMC bar
have similar, moderately subsolar metallicities
($[Fe/H] = -0.4$ dex; Westerlund 1997, see also Luck et al.~1998).
In this case, the 9M~CMD represents the largest homogeneous sample 
of non-variable supergiants and Cepheids ever assembled.

\subsection{ Isolating the Supergiants }

We present 
two different
versions of the 9M~CMD in Figure~2.
In these diagrams, we adopt a bin size of
0.03 mag in $(V-R)$ and $V$ with axes running from 
$-0.3 < (V-R) < 1.5$ and $17 < V < 14.5$ mag.   
The scale and bin size
are chosen to show the two supergiant sequences clearly. 
Intensity and contours 
(1.0 to 2.5 dex in 0.25 steps) indicate the logarithmic number
of stars (number per 0.03 mag square color-mag bin).
Panel (B) is a subset of panel (A), where we
have removed all variable stars. 
The variable stars are identified by poor fits to 
constant-brightness 4-year lightcurves.

There are only small regions 
in the 9M~CMD where the supergiant sequences are
free from contamination by other major features. 
This is particularly true for the blue supergiants.
The foreground Galactic disk stars are the most serious interloper
we must confront.  
In panel (B) of Figure~2, we indicate two ``apertures'' with white
rectangles.  These are used to count the (non-variable)
blue and red supergiants.  The apertures are
centered on $(V-R),V$ = 0.18, 15.25 and 0.66, 15.75 mag, respectively. 
The aperture sizes are $\Delta(V-R),\Delta V$ = 0.2, 0.5 mag. 
There are clearly more red
then blue supergiants in these apertures.  
We also draw attention to the gradients in number density
with increasing brightness within the apertures.
While the number density of blue supergiants is nearly constant, 
the number density of red supergiants is visibly decreasing. 
In this work, we restrict our analysis to the average properties
of supergiants in these two apertures\footnote{We have calculated
the ratio of blue and red supergiants in 88 quarter-field regions
of the 9M~CMD (including variables) and find a mean value
$0.504 \pm 0.033$ and a median value of 0.44, excluding one
region with an exceptionally high degree of differential
reddening.  This analysis reassures us that the mean ratio
of non-variable blue and red supergiants described in the text
is representative of the young LMC bar population.}.

In Figure 3, we present color-frequency histograms of the bright stars
in the 9M~CMD  (shown as solid line histograms).  These give
the number of stars per 0.1 mag color bin
and $V \pm 0.25$ mag, where $V$ is labeled in the upper right corner
of each panel.
The shaded histogram is a model prediction for Galactic
disk foreground stars along the line of sight toward the LMC bar.
The Galaxy model 
is briefly summarized in the next paragraph.   
The agreement between the 9M~CMD data and the model is comparable 
to other well-studied lines of sight (for a recent
modeling of starcount data along multiple lines-of-sight,
see Ng et al.~1997).  The detailed disagreement between the model and
9M~CMD starcounts may be
due to the normalization of the local disk luminosity function
and assumptions regarding the local disk giant branch
(see below).  This is not
critical to our analysis.  
This model comparison unambiguously identifies
feature H in Figure~1 with foreground stars.
In addition,
the color distribution of foreground Galactic disk stars 
shows a sharp blue edge, and thus
blue LMC supergiants may be isolated
with an appropriate color cut.  The fractional contamination 
of the red LMC supergiants by 
foreground Galactic disk stars is estimated to be
$\sim$10\% in the brightness range of
interest ($V = 15.75$ mag).

Our Galaxy model 
consists of two components, a standard double-exponential
disk and a spheroid.  However, the spheroid is a minor contributor 
to the starcounts at the brightnesses we are concerned with
in this work (Bahcall \& Soneira 1980),
and this component is not described here.
We do not include extinction in our model.  
Our local disk luminosity function is 
adopted from Yoshii et al.~(1987),
but is re-binned into
(no.~stars) $\times$ (0.5 mag)$^{-1}$ $\times$ pc$^{-3}$.
We predict a local spatial density of
$n$ = 0.036 stars pc$^{-3}$, which projects to a column density
at the solar radius of 47 $M_{\odot}$ pc$^{-2}$.  We adopt a CMD
for the disk stars consistent with the $z = 0.008$ isochrones
of Bertelli et al.~(1994).  This slightly subsolar metallicity
accounts for radial and vertical metallicity gradients 
in the disk, and dilution from thick disk stars in an average manner.
We have explicitly tested the effect of disk metallicity gradients
of the forms given by Yoshii et al.~(1987), and find a negligible
effect on the star count colors for the models relevant
to this investigation.  
We do not include a giant branch in our model.
We assume a Galactocentric solar radius R$_{\odot}$ = 8 kpc
and a disk radial scale length R$_{D}$ = 3.8 kpc (Yoshii
\& Rodgers 1989).  
Our vertical scale heights ($H_{Z}$), are a function of $M_{V}$,
running from $H_{Z} = 90$ for 
$M_{V} < 1$ mag to $H_{Z} = 400$ pc for $M_{V} > 5$ mag.
This function is consistent with the data 
summarized in Bahcall and Soneira (1980).
We have tested our Galaxy model against starcount data along lines
of sight toward the North Galactic pole (Yoshii et al.~1987), the South
Galactic pole (Reid \& Gilmore 1982), and the
Galactic anti-center (Ojha et al.~1994), and find
satisfactory agreement in all cases.

In Figure~4, we present a CMD also showing
a parallelogram connecting the non-variable
supergiant apertures used to count
Cepheids.  We plot the time-averaged magnitude and colors
of 1720 Cepheids identified in the 9M~CMD as small open circles
(not all of these are within the limits of Figure~4).
The smattering of very blue or red Cepheids may be attributed to
photometric blends or differential reddening and do not affect
this analysis.
We assume the catalog of Cepheids is complete at this brightness.
We also plot fiducial sequences for 
the non-variable supergiants 
as bold circles.
These fiducial marks are calculated
by assembling all stars within a 30 arcsecond radius of 
each Cepheid into color-frequency histograms,
and adopting the center of the highest bin.
The contrast of supergiants relative
to the foreground Galactic disk stars
is high because many Cepheids and non-variable supergiants
are grouped in clusters and loose associations.   
As a check,
we also perform a statistical subtraction
of foreground stars predicted by our Galaxy model
and remake the color-frequency histograms.
The fiducial marks derived in this manner are consistent. 
The supergiant fiducial marks are listed in Table~1, along with
other marks to be described later.

Also plotted in Figure~4 are two $z=0.008$, $M = 5 M_{\odot}$ 
stellar evolution model tracks.  These are
projected into the observable plane ($V$,$V-R$) with analytic
approximations to the Bertelli et al.~(1994) isochrone data.
The tracks have been shifted to ``fit'' the supergiant
fiducial sequences and also
the color of the main sequence turn-off (not shown).
The fit is judged by eye. 
For a fixed metallicity, the color difference between the
red and blue supergiants is
a function of initial mass.
The color difference between the main sequence
and red supergiants is also a function of initial mass
(at a fixed metallicity).  
Thus, we are confident that the
adopted apertures isolate supergiants with $\sim 5 M_{\odot}$
initial masses.  
The shifts are needed to place the model tracks ``in the LMC,''
i.e. to account for distance and reddening.
We disregard the sloping
initial mass function in this analysis because
the time $\sim 5 M_{\odot}$ stars
spend as helium-burning
supergiants is small compared to their main sequence lifetimes.
Although the model tracks extend
outside of the 
apertures (notably the red aperture), the time spent outside is
very small compared to the time spent inside.
The model which enters in
the lower right of the red aperture is from Schaerer et al.~(1993),
while the other model is from Fagatto et al.~(1994).  
The Schaerer et al.~(1993) model crosses
the Cepheid aperture at a brighter magnitude 
than the Fagatto et al.~(1994) model.
Both models have a metal abundance $z = 0.008$ and use the same
Livermore opacity tables. 
The Schaerer et al.~(1993) model has a helium abundance
$y = 0.265$, while the Fagatto et al.~(1994) model has
$y = 0.250$.  The treatments of convection (including overshooting)
and mass loss are similar, but may differ in some details.

\subsection{Counting Supergiants and Comparison with Theory}

The numbers of non-variable red supergiants, blue supergiants,
and Cepheids are $r = 2064$, $b = 805$, and $c = 280$, respectively.
If we make a 10\% correction to $r$ for foreground stars,
the blue to red supergiant ratio is $b/r$ = 0.43 and the ratio of Cepheids
to non-variable supergiants is $c/(b+r) = 0.105$.  We estimate an uncertainty
of 10\%, allowing primarily for
the uncertainty in our foreground star correction.
The blue
edge of the Cepheid instability strip in our aperture
is at $(V-R) = 0.30 \pm 0.02$ mag and the red edge is
at $(V-R) = 0.47 \pm 0.02$ mag.  Shifting the theoretical
models to match the fiducial supergiant sequences in the 9M~CMD provides
a self-consistent color-temperature calibration 
for these stars.   Using this calibration,
we find the blue and red edges are
$\log(T) = 3.77 \pm 0.01$ and 
$3.70 \pm 0.01$ dex, respectively.
We adopt the blue edge of the red aperture as
$\log(T) = 3.68$ dex.

Adopting the red edge of the blue aperture and
the blue edge of the red aperture to demark the red and blue supergiant
phases, the Schaerer et al.~(1993)
model predicts $b/r = 0.31$ while the Fagatto et al.~(1994) model
predicts $b/r = 1.02$.  We use the same segments of the model
tracks as shown in Figure~4. 
Adopting the instability strip as derived
above, we find $c/(b+r) = 0.096$ and 0.109 for
the Schaerer et al.~(1993) and Fagatto et al.~(1994) models,
respectively.  The agreement between our measurement and
the single mass model predictions for
$c/(b+r)$ is quite good.  This consistency
is strong support for these models 
accurately representing
the physical processes by which 
these stars evolve.
Once evolution across the Hertzsprung gap for 
the first blue loop is initiated in these models,
it may proceed through the instability strip on a time scale that 
is fairly insensitive to the detailed differences in
model input physics\footnote{This would make sense for the 
the first, or the 
fourth and fifth crossings (Becker 1981), because these proceed
on the thermal (Kelvin-Helmholtz) time scale of the envelope.
However, the first blue loop proceeds on a nuclear time scale
of the core (Kippenhahn \& Weigert 1990), which might 
be just as sensitive to model input physics as evolution
at the each end of the blue loop.}.
In contrast, it would appear
that evolution at each end of
the blue loop is quite
sensitive to differences in the model input physics.  
The nearly equal number of red and blue supergiants predicted
by the Fagatto et al.~(1994) model is difficult to reconcile
with our measured ratio.  
We speculate that subtle differences in the
parameterized treatments of convection may be the cause 
of this discrepancy between models.  

The agreement between our 
measured $c/(b+r)$ ratio for $\sim 5 M_{\odot}$
supergiants and the
model predictions has several interesting implications.  
If the instability strip crossing time is
nearly independent of model input physics, we infer that the models
accurately predict the total time these stars spend as 
helium-burning supergiants.
Since the ratio of hydrogen to helium-burning lifetimes 
for intermediate-mass stars is sensitive to the
treatment of convective overshooting (Lattanzio et al.~1991),
our result lends support to the overall amount of 
parameterized convective overshoot in these two models
examined, despite the detailed differences found.

There are additional implications for the Cepheids.
A theoretical Cepheid period-frequency histogram calculated by
assuming a constant star formation rate and a Salpeter (1955)
initial mass function 
predicts fewer $\sim 5 M_{\odot}$ Cepheids than observed 
in the LMC bar, which has been interpreted 
as a short duration ``burst'' of star formation $\sim$100 Myr ago
(Alcock et al.~1998b).
The inference of a burst is likely a
model-independent result because
the time spent in the instability strip is very nearly the same
for the two $5 M_{\odot}$ models examined here.
However, we caution
that results pertaining to the peak of the Cepheid period-frequency
histogram in Alcock et al.~(1998b) are strongly model dependent,
since the peak number of Cepheids in a period-frequency
histogram corresponds to the time spent at the
tips of those blue loops that enter the instability strip
($M < 5 M_{\odot}$ in the LMC).
Finally, we note that
neither of the two $5 M_{\odot}$ 
models examined here have second blue loops.  However, the
uncertainty of our measured $c/(b+r)$ ratio
is too large to rule out or confirm the existence 
of a small number of Cepheids on second blue loops.

An additional interesting implication of the $c/(b+r)$ 
ratio in the 9M~CMD
pertains to the discovery 
of $\sim$600 Cepheids with $V \simgt 17$ mag and
periods $\simlt$ 2.5 day (Alcock et al.~1998b, see also 
Alves et al.~1998).
These Cepheids clearly lie below the
observed blue supergiant sequence in the 9M~CMD, implying
evolution via a ``non-standard'' channel. 
Alcock et al.~(1998b) propose that the faint Cepheids are 
merged binaries, formerly
two $\sim$1.5 $M_{\odot}$ stars.  Indeed, these stars 
may be ubiquitous if a major burst of star formation
occurred $\sim$3 Gyr ago.  The number of merged
$\sim$2.5 $M_{\odot}$ stars that produce $\sim$5 $M_{\odot}$ Cepheids
is calculated by scaling from the faint Cepheids, as follows.
We assume a Salpeter (1955) initial mass function, vis.
$dN/dM = M^{-2.35}$, and a constant star formation rate
over the period of time during which 1.5 to 2.5 $M_{\odot}$
Cepheids are born in the LMC.  In this case, the
number of $\sim 5 M_{\odot}$ Cepheids produced by binary mergers
would be $n \approx 600 \times (1.5/2.5)^{2.35} \approx 180$.  Adding this
to the number of Cepheids predicted from the 
number of red and blue supergiants 
yields $c$~=~460, which compares with $c$~=~280 observed.

It may be difficult to reconcile the binary merger hypothesis
for the origin of the low luminosity Cepheids with the 
$c/(b+r)$ ratio for $\sim$5 $M_{\odot}$ Cepheids.
First, it is unlikely that merged binaries will follow single
star evolution paths through the red and blue
supergiant phases if the Cepheid loop is triggered
by the discontinuity in the chemical profile of the envelope
as suggested by theory.  
There is no obvious mechanism by which the necessary
envelope discontinuity would form during, or after, a binary merger.
Therefore, merged binaries would not produce the correct
numbers of Cepheids and non-variable red and blue supergiants.
Second, while it is possible that the star formation history of the LMC
conspired to produce very few 2.5 $M_{\odot}$ stars (thus avoiding the
problem of too many $\sim$5 $M_{\odot}$ merged-binary Cepheids),
this requires very fine tuning of the star formation history.
If the 5 $M_{\odot}$ models are to be trusted,
we conclude either (1) the probability of binary mergers is much
lower for higher mass LMC stars, and thus only low luminosity
Cepheids are produced from binary mergers, or (2) 
binary mergers are not responsible for the curious low
luminosity Cepheids.

\section{Old Stellar Populations: The Data}

\subsection{9M~CMD Fiducial Marks}

Fiducial marks on the giant branch are defined as the
peak of color-frequency histograms in $\Delta V$ = 0.5 mag bins centered
on the values listed in Table~1 ($V$ = 16.75, 17.25 17.75, 18.25, and 19.75 mag).
The fiducial marks were measured separately 
for each of 22 MACHO survey fields\footnote{Although it is beyond the
scope of this work to provide a detailed comparative analysis 
of these 22, $\sim$0.4
million star CMDs, we remark on a few outstanding field giant branches
(see Alcock et al.~1997, their Fig.~1).
Fields 11 and 82 appear to suffer 
from notably large amounts of differential reddening. 
Fields 3, 77, and 80 are quite red (these are adjacent, 
running North from near the center
of the bar).  
We speculate that there is a higher degree of foreground reddening for these
fields (i.e. possibly a foreground dust cloud).  Field 7
had the bluest giant branch 
(it is adjacent and South of field 77), and we suspect
a particularly small degree of foreground reddening.}
included in the 9M~CMD (see Alcock et al.~1997).  
In some cases, no clear peak was evident at $V$ =  19.75 mag 
(fields 11, 15, and 81), and these fields were excluded when
calculating this average mark.  
The average marks should accurately
represent the shape (i.e. curvature) 
of the giant branch in the 9M~CMD; they are listed in Table~1.
We find a standard deviation for each of the mean RGB fiducial marks of
$\sigma_{V-R}$ = 0.04 mag, which gives an indication of
the natural width of the giant branch 
(due possibly to superposed RGBs and AGBs,
but also differential reddening throughout the bar).

In the top panel of Figure~5, we plot as small circles
a random $\sim$10k stars from the 9M~CMD
located near the tip of the RGB (tip-RGB).
In the middle panel, we show a different
$\sim$10k sample of stars, where we have now 
excluded all candidate variable stars.
The horizontal lines in these two panels are the giant branch
fiducial marks (these show $\pm 0.04$ mag in color).  
The angled mark corresponds to $W_{3.3} = V - 3.3(V-R) = 13.62$ mag
for $0.80 < V-R < 0.95$ mag.  The derivation
of this fiducial mark is illustrated in the bottom panel.  Here,
we plot histograms of all stars in this region of the 9M~CMD
(not just the $\sim$10k subsets shown in the upper two panels)
projected along the $W_{3.3} = V - 3.3(V-R)$ vector.   
The bin-size is 0.02 mag.
The solid line shows all stars, while the dotted
line shows all non-variable stars.  
The luminosity functions fall off at large $W_{3.3}$ because we only
count stars within the limits of $(V,V-R)$ shown in the upper 
two panels.
The slope of the $W_{3.3}$ vector 
was chosen by eye.  (Trials with different
slopes did not change the result.)
We calculate the difference in height between 
consecutive bins, divided by the square root 
of the mean number of stars in the
two bins as a measure of the ``step'' which occurs at the tip-RGB.  Both 
histograms yield $W_{3.3}$ = 13.62 as the most significant step, which
is marked with an arrow in the bottom panel of Figure~6.
However, the lack of a well-defined tip-RGB is the most notable
feature of these $W_{3.3}$ luminosity functions.
This may be due to photometric blends, differential reddening,
mixed age and metallicity LMC populations
on the giant branch, or a combination of these effects.
In Table~1, we list the blue and red tip-RGB 
at $(V-R)$ = 0.80 and 0.95 mag, $W_{3.3}$ = 13.62 mag.

In Figure 6, we show a log-scaled 
Hess diagram of the region around the horizontal branch
in the 9M~CMD.  Axes run from $(V-R) = -0.1$ to 0.7 mag 
and $V$ = 20 to 18 mag.   
Bin size is $\Delta(V-R)$ = 0.01 mag and $\Delta V$ = 0.02 mag.
We plot logarithmic 
contours at 1.8, 2.0, 2.2, 2.4, 2.6, 2.7 (near the AGB-bump),
2.8, 3.0, 3.2 (near the red HB clump), 3.3 and 3.4 dex.  
On the right-hand axis of Figure~6 and in order
of decreasing brightness, we
mark the fiducial $V$ magnitudes for
the AGB-bump, red HB clump, and field RR~Lyrae variables.  
These brightnesses and
associated colors are also listed in Table~1.  
For the first two, we adopt the
peak pixel value from smoothed versions of the
9M~CMD. Two-dimensional gaussian profile 
fitting to these ``bumps'' yielded consistent results.
The derivation of the median brightness of RR~Lyraes
is described in the next section.

\subsection{ RR Lyrae Sample Selection }

The sample of RR Lyraes was selected as follows.
We used a period--amplitude diagram from the
one-year variable star catalog\footnote{This catalog consists of
all LMC variables identified with poor fits to constant-brightness
one-year MACHO lightcurves and includes
phasing information.}
to identify candidate type ``ab'' RR Lyrae 
(RRab)
with periods between 0.46 and 0.71 day and amplitudes between 0.1 and 1.7 mag
(MACHO instrumental blue photometry).  
Four-year lightcurves were then extracted
for 3728 stars, excluding candidates in 6 fields 
covering the central region of
the LMC bar\footnote{Exlcuded fields are 1, 7, 9, 77, 78 and 79
(see Alcock et al.~1997, their Figure~1).
These are ``Round-1'' fields with a different naming
convention in the year-one catalog and the
four-year lightcurve database.}.  Next, approximately 185 candidate RRab were
then eliminated for poor lightcurve quality and uncertain periods
(periods were derived with a ``supersmoother''; Reimann 1994).
We sorted the individual photometric measurements 
in each lightcurve by increasing brightness
and adopted the 3rd and 97th percentiles as maximum and minimum light,
which yielded pulsation amplitudes (a procedure
robust against outlying measurements 
and also small errors in the periods).
A few stars with suspect amplitude ratios ($A_V$/$A_R$)
were discarded (a cut designed to remove
eclipsing binaries in the sample; Minniti et al.~1996).
Time-averaged, flux-weighted mean magnitudes
and colors were then calculated, and
a small number of stars with very red
colors ($V-R > 0.55$ mag) were also discarded.
The final sample contains 3454 RRab stars.

The completeness
of this RRab catalog is uncertain.
Preliminary artificial star tests suggest that
we are 50--90\% complete for stars at the mean
brightness of the LMC RRab.  Furthermore,
these tests demonstrate that our photometry is relatively
unaffected by Malmquist bias.  
If we detect an RRab in our image data, then 
we will likely identify it as a variable because of the
typically large pulsation amplitudes.
Other effects, such as period aliasing in the MACHO
photometry data (i.e.~at 0.5 day), will tend to lower the
completeness. 
There is no reason to suspect that
this RRab catalog suffers from any spatially-dependent incompleteness bias.
We emphasize that the
16 fields analysed here do not include the most crowded
fields in the very center of the bar.
A modest degree of spatially-uniform incompleteness in
this RRab catalog will not affect the results of this paper.

We define two subsamples of the RRab catalog in order to
calculate different mean properties.
The first subsample we define is for consideration of
the period-amplitude diagram, which we will refer to as the
Bailey diagram sample.  We begin with the $\sim$35\%
least crowded\footnote{For each photometric measurement 
in the full MACHO database, there is
a ``crowding parameter'' which gives the percentage flux inside the
PSF fit box contributed by neighboring stars.
It is useful to think of the least crowded stars as living in ``empty''
patches of sky scattered througout the otherwise crowded LMC bar.},
or about 1280 RRab.
Next, we make strict cuts in
magnitude $19.7 < V < 19.2$, eliminating highly reddened 
stars (whose amplitudes may be underestimated because of Malmquist
bias at minimum light), the numerous blended stars (whose amplitudes will
certainly be underestimated because of the contaminating flux),
and foreground RR Lyrae (Alcock et al.~1997b).  This cut
is similar to a ``sigma-clip,'' and leaves 935 RRab.

Next, we define a CMD RRab sample.
We remove the sigma-clip to guard 
against a possible bias in our estimate 
of the median color and brightness of the RRab.
However, we tighten the crowding cut to exclude all but the $\sim$20\%
least crowded, or about 680 RRab.  Without the sigma-clip, this
sample contains a few
bright RRab blended with blue (main sequence) and red (giant branch or clump) 
stars in the CMD.   Some of the brighter stars may also be 
evolving off of the ``zero-age'' horizontal branch.
There are also fainter and redder RRab which are most likely
affected by differential reddening in the LMC.  
The median color and
magnitude of this RRab subsample is $V$ = 19.45,
$(V-R)$ = 0.31 mag, which is listed in Table~1.

The ancient LMC clusters
NGC~1835 and NGC~1898
each reside in one
of the 6 fields excluded from the RRab catalog described above. 
Therefore, we have identified a few RRab in these clusters
by culling through the MACHO database ``by hand.''
No effort was made to identify complete samples.
Variables
lying within $\sim$1 arcmin of each cluster center are assumed
to be members.  We identify 8 RRab in NGC~1835 and 5 RRab in NGC~1898.
This is the first report of RR~Lyrae in NGC~1898.  
Walker (1993)
and Graham and Ruiz (1977) have also discovered RR~Lyraes in NGC~1835.
The properties of the RRab in these two LMC bar clusters are
summarized in Tables~2 and 3.  
We assume the mean colors
and magnitudes are representative of each cluster.

Last, we refer to Alcock et al.~(1997c) for the discovery of 75 multimode
(RRd) stars in all 22 of these fields.  
The calibration of MACHO photometry to Kron-Cousins $V$ and $R$ 
described in
Alcock et al.~(1999) supersedes the calibration used in Alcock et al.~(1997c).
Mean properties of these RRd are now: $<V> = 19.327 \pm 0.021$ and
$<V-R> = 0.259 \pm 0.006$ mag.  

\subsection{ AGB Sample Selection }

In order to study the
AGB in the 9M~CMD, we begin with the $\sim$88000 candidates in 
the four-year LMC variable star catalog.  Each lightcurves is calibrated
to $V$ and $R$ and time-averaged, flux-weighted
mean magnitudes and colors are calculated. 
The statistical cuts used to identify candidate variable
stars are looser in this four-year catalog
than those used to generate
the one-year catalog (the latter was used to select the RRab).
We remind the reader that the MACHO variable star catalogs are a by-product
of the microlensing searches (Alcock et al.~1997),
and subject to the designs of those analyses, not the present one.
In order to 
``clean-up'' the relatively loose-cut four-year variable star catalog,
we calculate a Welch-Stetson variability index (Welch \& Stetson 1993)
and several lightcurve quality statistics. 
Applying cuts with these statistics,
we reduce the catalog to a sample of $\sim$19000 of the 
most significantly 
variable stars, including only those with 
the highest quality lightcurve data.
The AGB variables in this ``clean'' sample 
are analysed in the next section.

\section{ Old Stellar Populations: Analysis }

Simple theories of galactic chemical evolution 
predict that progressively
younger stellar populations will be more
metal-rich.
Analyses of planetary nebulae 
confirm that the young stellar populations 
in the LMC are more metal-rich than the old
stellar populations (Dopita et al.~1997).
Age--metallicity data for LMC clusters show the same trend
(Da Costa 1998), although comparing the field and cluster
histories may not be strictly appropriate (van~den~Bergh 1998).
If two old
populations are required to account for 
features in the 9M~CMD, we assume that the younger population
is relatively more metal-rich.

\subsection{ The AGB }

In Figure~7, we plot the clean sample of AGB variables 
with small circles.  We show the tip-RGB as the steeply-angled
mark running from $0.80 < V-R < 0.95$ mag.  
The thin solid line corresponds
to $W_{2.0} = V - 2.0(V-R) = 13.5$ mag, 
which runs approximately parallel to the extended
sequences of AGB stars.  
AGB variables from
the LMC clusters NGC~1898 (bold dots) and
NGC~1783 (bold circles) are also shown.
The data for these cluster AGB variables are summarized
in Table~4 (see Alves et al.~1998 for lightcurves). 

NGC~1783 is a $\sim$2 Gyr old cluster located well 
outside of the LMC bar (Mould et al.~1989).  
The MACHO photometry data for this outer disk LMC field
are transformed to Kron-Cousins $V$ and $R$ by comparison 
with the standard stars of Alvarado et al. (1995).
Finding charts and original AGB star identifications are found
in Lloyd Evans (1980), near-infrared photometry and $m_{bol}$ are assembled
from Frogel et al. (1990), and spectral types from Mould et al. (1989).
The spectroscopic metallicity of NGC~1783 is [Fe/H] = $-0.5$ dex (Cohen 1982).
NGC~1898 is an ancient LMC cluster located in the bar
(Olsen et al.~1998).
The spectroscopic metallicity is [Fe/H] = $-1.4$ dex (Olszewski et al.~1991).
Finding charts, near-infrared
photometry, and $m_{bol}$ for these AGB stars are given in
Aaronson and Mould (1985).

The AGB in Figure~7 shows a rich morphology.
We draw attention to the concentration of AGB variables
lying near to the tip-RGB.  Offset from this
concentration are two extended sequences, 
both running approximately parallel
to the $W_{2.0}$ = 13.5 line.  The fainter sequence is the 
most prominent.  The variables from NGC~1783
lie along the brighter sequence, showing that the
variables on the brighter AGB sequence are consistent 
with an ``intermediate-age'' LMC bar population
(like NGC~1783).
The variables from
NGC~1898 lie in the concentration near the tip-RGB, and not along
the either of the extended sequences.  We conclude that
the concentration of AGB variables near the tip-RGB are 
consistent with an old and metal-poor LMC bar population
(like NGC~1898).

Figure~8 shows four 
$W_{2.0} = V - 2.0(V-R)$ luminosity functions.  
Three of these histograms represent different color cuts through
the AGB variables shown in Figure~7.
The dotted line represents
$0.8 < V-R < 1.1$ mag, the solid line with open circles shows
$1.1 < V-R < 1.4$ mag, and the solid line with solid circles
shows $1.4 < V-R < 2.0$ mag.  These $W_{2.0}$ luminosity functions have bin
sizes of 0.1 mag; the left axis 
gives the number of stars in each bin.
The $W_{2.0}$ luminosity function for 266 known carbon stars
(Blanco, McCarthy, \& Blanco 1985) 
is indicated with the shaded histogram
(the bin size is also
0.1 mag).  The number of carbon stars is given on the right axis.
We mark $W_{2.0}$ = 13.5 with an arrow, which corresponds to
the thin solid line in Figure~7.

First, consider
the 9M~CMD stars in the two histograms with the reddest
colors (open circles and solid circles in Fig.~7).
{\it These clearly show the bimodality of the AGB.\ }   
We note that the ratio of the number of stars on the
faint AGB sequence to those on the bright AGB sequence
is higher in the intermediate color-cut histogram (open circles) 
than in the reddest color-cut histogram (filled circles),
an effect primarily due to the relatively more
rapid decrease in the number of stars 
along the faint sequence.

The distribution of carbon stars rises sharply at 
$W_{2.0} \approx$ 13.5 mag,
coincident with the rise in the number
of AGB variables.  We do not
know the completeness of this sample of carbon stars identified 
in the 9M~CMD.  However, if the true
carbon star luminosity function does not rise dramatically
(i.e., a second peak in $W_{2.0}$), 
we conclude that the majority of stars 
on the faint AGB sequence and those
concentrated near the tip-RGB (dotted line)
are likely oxygen-rich, not carbon-rich.
Note that the AGB variables in the clusters 
are mostly located along one or the other AGB sequence. 
This suggests how AGB stars 
from different age populations evolve through the 9M~CMD.
Intermediate-age populations produce both carbon 
and oxygen-rich AGB stars, but
very old populations produce only oxygen-rich AGB stars. 
Surveys for carbon stars in clusters tell us that
the characteristic transition age for a stellar population to begin
producing carbon stars increases
with decreasing metallicity.  This transition age
may be $\sim$3 Gyr for solar
metallicity, but is much older for low metallicities, i.e.
$\sim$10 Gyr at $[Fe/H] \sim -1.5$ dex (Bessell, Wood, \& Lloyd Evans 1983).  
No ancient LMC clusters have
carbon stars (Frogel et al.~1990).  
Therefore, if we assume that the two extended AGB sequences arise
from distinct populations, and assume that the population responsible
for the faint AGB sequence does not
produce carbon stars, then if the 
AGB variables on the faint sequence
are metal-poor, they are likely quite old.

In summary, most of the highly-evolved 
AGB stars in the 9M~CMD are variable.  Some of the AGB
variables concentrate near the tip-RGB.  These stars are
consistent with being old and metal-poor, 
like AGB variables in the
cluster NGC~1898.
Offset from this concentration are two extended AGB sequences, 
both running approximately parallel to the vector $W_{2.0} = V - 2.0(V-R)$.   
The AGB variables on the bright sequence
are consistent with being $\sim$2 Gyrs old and
having a metallicity $[Fe/H] \sim -0.5$,
like AGB variables in NGC~1783.  The majority of AGB variables on the 
faint sequence and in the concentration near the tip-RGB are likely
oxygen-rich, not carbon-rich.  This may imply they are quite old,
if they are fairly metal-poor.  
We caution that the latter inference depends
on the completeness of the 
Blanco, McCarthy, and Blanco (1985) carbon star survey data 
in the MACHO fields analyzed here.

Stellar evolution theory relies heavily on observations of AGB stars
in the LMC to tune various free parameters in AGB star models.
For example, the models are tuned to reproduce the carbon star luminosity
function in the LMC and SMC (Marigo et al.~1996).
Therefore, logical inferences from these models 
regarding the detailed make-up of the LMC stellar populations
would be disconcertingly circular.  However, we note that
our discussion about which 
stellar populations produce carbon 
stars, and what path of AGB evolution is followed through the 9M~CMD
is consistent with extant theory.
It would be particularly interesting to
use the numbers of AGB stars along the extended
sequences to derive stellar evolutionary lifetimes and test
the AGB models.  In addition,
analysis of the pulsation
properties of these many AGB variables will be very important 
to the accurate interpretation of their ages and metallicities.

\subsection{The RR Lyrae}

Figure~9 is the Bailey period-amplitude diagram, which shows
the RRab from the 9M~CMD (small circles), those from NGC~1898
(bold circles), and those from NGC~1835 
(bold triangles).  In the inset, we plot RRab
from the Galactic globular cluster M3 (Kaluzny et al.~1998).  
The axes of the inset are
the same as those in the main panel.  
We define a reduced period,
\begin{equation}
\log(PA) = \log(P) + 0.15 A_V
\end{equation}
and calculate the median value for M3, $\log(PA)_{M3} = -0.1116$.  
This line is plotted in the inset, and again in the main panel.
The median reduced period of the 
RRab in the 9M~CMD is $\log(PA)_{9M} = -0.1110$.
Although the 9M~CMD RRab define a prominent ridge line, their
distribution is asymmetric.  They form a ``cloud'' to the right of the
ridge line with larger amplitudes and longer periods.

The reduced period correlates
with metallicity.  We derive the following calibration using
high-quality $V$-band lightcurves of RRab in
the Galactic globular clusters M3, M5, and M15.  Data for
M5 are assembled from Reid (1996), M3 
from Carretta et al.~(1998), and M15 from Silbermann and Smith
(1995), from which we calculate median reduced periods: 
$\log(PA)_{M5} = -0.1352$, 
$\log(PA)_{M3} = -0.1117$,
$\log(PA)_{M15} = -0.0646$.
We adopt metallicities of $[Fe/H] = -1.4$, $-1.6$, and $-2.1$ 
taken from Sandquist et al.~(1996),
Carretta et al.~(1998), and Silbermann and Smith (1995)
for M5, M3, and M15, respectively.   These yield
a calibration,
\begin{equation}
[Fe/H] = -8.85 \log(PA) - 2.60 
\end{equation}
As a check of this calibration, we assemble
$A_V$, $\log(P)$, and spectroscopic $[Fe/H]$ data 
for 86 nearby Galactic field RRab
from Jurscik and Kovacs (1996).  We shift the
Jurscik and Kovacs (1996) metallicity scale
by $-0.2$ dex to place it on the scale of our clusters. 
Our calibration  predicts the metallicity of these RRab 
with an accuracy 
of $\sigma_{[Fe/H]} = 0.31$ per star.

The median metallicity of 
RRab in the 9M~CMD is $[Fe/H] = -1.6$,
which agrees with the mean spectroscopic metallicity
of 15 field RRab presented by Alcock et al.~(1996), 
$[Fe/H] = -1.7 \pm 0.2$ dex on the scale of Zinn and West (1984).
For the NGC~1898 and NGC~1835 cluster RRab, we find
$[Fe/H] = -1.58 \pm 0.10$ and $-1.95 \pm 0.09$ dex, respectively
(errors are statistical only).
Our metallicity for NGC~1835 agrees with
the spectroscopic value ($-1.8 \pm 0.2$ dex) from Olszewski et al.~(1991) 
and Walker's (1993) estimate of $-1.8$ dex.  
Our metallicity for NGC~1898 is slightly lower than the spectroscopic
metallicity ($-1.4 \pm 0.2$ dex) found by Olszewski et al.~(1991).
Following the discussion of Kaluzny et al.~(1998) for M3, 
the metallicity of RRab in the 9M~CMD may be
$[Fe/H] \approx -1.4$ on the scale of 
Carretta and Gratton (1997).

The RRab in the LMC bar define a 
prominent ridge line in the Bailey diagram
similar to the ridge line defined by RRab in M3,
which we interpret
as indicating similar mean metallicities.  
The 9M~CMD RRab
with higher amplitudes and longer periods
are consistent with a tail in the metallicity
distribution to lower values.  Observational evidence strongly
suggests that some RR~Lyraes in the LMC
have metallicities as low as 
$-2.3$ dex (Walker 1992, Alcock et al.~1996).
However, we caution that the distribution of RRab in the
Bailey diagram is subject to evolutionary effects,
and perhaps also population effects (i.e. age), which
we have neglected here.

\subsection{The Giant Branch and AGB-Bump}

When cluster ages
are the same to within a few Gyr, 
the dereddened
colors of their giant branches yield accurate relative metallicities
(Sandage \& Smith 1966, Da Costa \& Mould 1986, Sarajedini 1994). 
Each panel in Figure~10 compares the
9M~CMD horizontal
branch and the giant branch fiducial marks
with CMD data for three different 
clusters.  The 9M~CMD fiducial marks are as follows.
The median color and brightness of the RRab 
are indicated with an open triangle, the red HB clump
with a large open circle, and the AGB-bump 
with a small open circle.  Dash marks show the
giant branch and the tip-RGB.

In the top panel
of Figure~10, we plot the CMD of M5 
(Sandquist et al.~1996).
In the middle panel, we plot the CMD data of M3
(Ferraro et al.~1997).  In these two cases, we
have transformed their photometry 
using  $(V-R) = 0.557(B-V) + 0.019$, which follows
from Alcock et al.~(1997c; see also
Bessell 1990).  Next, we calculate shifts 
to match the mean magnitudes and colors
of the cluster RRab with the 9M~CMD fiducial mark.  For M5,
we assemble $(B-V)$ data for 11 RRab from Storm,
Carney, and Beck (1991; see also Sandquist et al.~1996)
and calculate $<B-V>_{M5} = 0.292$ and $<V>_{M5} = 15.057$ mag,
yielding shifts of $\Delta(V-R)_{M5} = 0.13$ and
$\Delta V_{M5} = 4.39$ mag to place this cluster ``in the LMC.''  
Using the sample of M3 RRab from Carretta et al.~(1997), we find
$<B-V>_{M3} = 0.376$ and $<V>_{M3} = 15.688$ mag,
and shifts of $\Delta(V-R)_{M3} = 0.08$ and
$\Delta V_{M3} = 3.76$ mag. 
In the bottom panel of Figure~10, we show 
the SMC cluster NGC~411,
which is 1.5 Gyr old and
has a metallicity of $[Fe/H] = -0.7$ (Alves \& Sarajedini 1999).
With the NGC~411 data transformed to $(V-R)$ as above,
they are shifted $\Delta(V-R)_{N411} = 0.06$ and
$\Delta V_{N411} = -0.22$ mag to match the location of the 9M~CMD
red HB clump (the fit is by eye).  

Figure 10 shows that 
the color of the giant branch in the 9M~CMD is well-matched
by the giant branch of M3.  
If the difference
in color between the RRab and the giant branch, $\delta(V-R)^{RRab}_{GB}$,
depends only on metallicity, this may imply that
the metallicity of the
old field population in the bar is similar to that of M3.
However, this interpretation ignores the influence of the 
second parameter (Sandage \& Wildey 1967, Stetson et al.~1996)
and the finite width of the fundamental mode instability strip
(Bono et al.~1997).  Although we avoided making numerous assumptions 
about the distance and reddening to the LMC and these clusters
by registering the RRab,
metallicity and the second parameter are degenerate
in this comparison.

Consider $\delta(V-R)^{RRab}_{GB}$
for the classical
second-parameter pair of Galactic globular clusters, M3 and M13. 
We assemble the
photometry of Guarnieri et al.~(1993) for M13
and the photometry of Pike and Meston (1977) for the one RRab in
this cluster.  We find $\delta(V-R)^{RRab}_{GB} = 0.36$ mag for M13,
which compares to $\delta(V-R)^{RRab}_{GB} = 0.233$ mag for M3.
Using the HB morphology index\footnote{This index is defined as the number
difference between blue ($b$) and red ($r$) HB stars,
divided by the total number of HB stars, including
the number of RR Lyrae ($v$).  We use lower case to distinguish
from magnitudes, i.e. $B$ $V$ $R$.  
The index runs from $-1$ to 1 for completely
red and blue HBs, respectively.}
of Lee et al.~(1994),
these clusters have $(b-r)/(b+v+r)$ = +0.08 
and +0.97, respectively (Catelan \& de Freitas Pacheco 1995).
The cluster with the redder HB morphology has 
redder RRab (although M13 has only one RRab).  
For comparison, a spline fit to the
giant branch fiducial marks yields
$\delta(V-R)^{RRab}_{GB} = 0.207$ for the 9M~CMD,
which is close to the value for M3.  

Figure~10 also shows that
the giant branch and red HB clump in the 9M~CMD are consistent with 
an NGC~411-like population.  
An ``intermediate age'' population
in the LMC bar older than $\sim$1.5 Gyr and more metal-rich
than $[Fe/H] = -0.7$ dex is unlikely to contribute significantly
to the 9M~CMD giant branch.  This is somewhat more metal-poor
(and younger) than often suggested for the intermediate age
population in the LMC, but is consistent with other recent
analyses (e.g.~Bica et al.~1998).  
It is unlikely that an NGC~411-like population accounts for
{\it all} of the red HB clump stars in the 9M~CMD unless the main-sequence
luminosity function analysis of Hardy et al.~(1984) is 
seriously in error.   
Finally, we note that an M3-like HB is not responsible
for the bright red HB clump in the 9M~CMD.

We have drawn arrows in Figure~10
indicating the AGB-bumps in M3 and M5.
The cluster AGB-bumps have similar brightnesses to the 
AGB-bump in the 9M~CMD, which supports our identification of this feature
(see also Gallart 1998), and our association of the AGB-bump 
with an old and metal-poor population.  
However, the AGB-bump in M3 is 
distinctly bluer than the AGB-bump in the 9M~CMD.
At a constant metallicity,
the AGB-bump will appear redder
for progressively more
massive HB progenitor stars, until the AGB-bump
lies near the Hayashi line.
At this point, even more massive HB stars will
populate similar color AGB-bumps, but with 
higher luminosities
(Castellani, Chieffi, \& Pulone 1991).
The AGB-bump for a population as young as NGC~411 would be $\sim$0.7 mag
brighter than the AGB-bump in the 9M~CMD
(Alves \& Sarajedini 1999).
In summary, the brightness of the AGB-bump in the 9M~CMD 
associates this feature with an old population, while the color 
of the AGB-bump strongly suggests a
red HB morphology.

\subsection{ The Horizontal Branch and AGB-Bump }

In Figure~11,
the 9M~CMD fiducial marks for the RRab,
red HB clump, the AGB-bump, and giant branch are plotted 
with the same
symbols used in Figure~10 (except that the red HB clump and AGB-bump are 
now filled circles).
We plot individual RRab stars as small open circles.  
The mean magnitudes and colors of the
RRab in the clusters NGC~1898 and NGC~1835 are shown
with error bars.
We additionally
plot twenty-four candidate post-HB
variables (large open circles), also known as BL~Hers\footnote{
BL~Hers are related to the W~Virginis (Type~II Cepheids) and
RV~Tauri variables (Alcock et al.~1998).}.  
These post-HB variables were found in a search 
of the four-year catalog (see \S4.3) for stars with
$17.5 < V < 18.8$ mag, $0.15 < (V-R) < 0.75$ mag, and 
periods in the range of $0.73 < P < 5$ day.  
No effort was made to identify a complete sample.  
The post-HB variables help define the instability strip (IS) for
the old and metal-poor population in the 9M~CMD.

We also plot the $M = 0.70$ and 0.75 $M_{\odot}$,
$z$=0.0004, scaled-solar
HB models of Castellani, Chieffi,
and Pulone (1991) in Figure~11  with solid lines. 
The lower mass model begins on the zero-age
HB near the center of the instability strip (dashed lines).   
Both of the HB
model tracks are truncated at their mean AGB-bump luminosity. 
The color-temperature calibration is 
the same analytic
approximation to the Bertelli et al.~(1994) isochrone data
used throughout this paper.
In this analysis, we shift the end-points of
these two HB model tracks to match the location of the AGB-bump
in the 9M~CMD, which 
re-calibrates color and temperature to account for the
LMC distance and reddening.
We show the prediction of Bono et al.~(1997)
for the fundamental blue and red edge of the 
instability strip (IS), for a mass $M = 0.725 M_{\odot}$.
The good agreement between the theoretical
IS and the distribution
of these 9M~CMD variables 
over several magnitudes in brightness strongly supports
our color-temperature calibration, and the theoretically-predicted
IS for the fundamental mode pulsators (Bono et al.~1997).

Figure~11 shows that the 9M~CMD field RRab
lie very close to the red edge of the
IS.  The tail of RRab fainter and redder than
the theoretical IS are likely due to 
differential reddening in the LMC bar.
The median brightness and color of
the 9M~CMD field RRab are consistent with the mean color and brightness
of RRab 
in NGC~1898, and significantly redder than the
mean color and brightness of
RRab in NGC~1835.  The HB morphology indices 
for NGC~1835 and NGC~1898
are $(b-r)/(b+v+r)$ = $+0.48 \pm 0.05$ and
$-0.08 \pm 0.10$, respectively (Olsen et al.~1998).
As for the case of M3 and M13, 
these relative mean RRab colors correlate with the
overall HB morphology\footnote{The LMC clusters NGC~1898 and NGC~1835 have
different metallicities, which is not the case for M3 and M13.   Thus,
the LMC clusters are not necessarily showing second parameter
effects.}.
Thus, we infer an HB morphology for the old and metal-poor
field population in the 9M~CMD that is at least ``as red'' as the
HB in NGC~1898.
Specifically, this comparison suggests
an HB morphology index of
$(b-r)/(b+v+r)$ $\simlt$ 0.

Next, we consider the HB morphology index for the two HB models
shown in Figure~11.
For the $M = 0.70 M_{\odot}$ model, the
time spent in the IS is $\sim$90 Myr, while the
time spent in the AGB-bump is $\sim$4 Myr.  
This model predicts
a ratio of AGB-bump to HB stars $a/(b+v+r) \approx 0.05$.
This model would correspond
to the case of $(b-r)/(b+v+r) \approx 0$
because the model track begins near the middle of the IS.
The $M = 0.75 M_{\odot}$ model predicts $\sim$22 Myr in the
IS, $\sim$65 Myr redward of the IS, and $\sim$4 Myr in
the AGB-bump.  This model also predicts $a/(b+v+r) \approx 0.05$.
If we assume $b << (v+r)$, the $M = 0.75 M_{\odot}$ model
predicts $r/v \approx 3$, $a/v \approx 0.2$,
and $(b-r)/(b+v+r) \approx -0.75$.  
This is a plausible HB morphology for this model given the
red color of the zero-age HB 
and assuming a small HB mass dispersion, as
used for model HB constructions (e.g. Lee et al.~1994).
In summary, the $M = 0.70 M_{\odot}$ model should
have $(b-r)/(b+v+r) \approx 0$ while the 
$M = 0.75 M_{\odot}$ model should have
$(b-r)/(b+v+r) \approx -0.75$. 

Assuming that our
RRab catalog is 50\% complete, correcting for the
6 excluded fields (22/16), and accounting 
for type ``c'' variables
with a ratio of RRab/RRc = 3/2 (Kinman et al.~1991), 
we estimate the total 
number of RR~Lyraes in the 9M~CMD to be
$v \approx 15000$~stars.
We count the number
of AGB-bump stars as follows.  First, we fit a power-law 
to a differential luminosity function 
of the giant branch
(the number of stars with $(V-R) > 0.4$ mag
per 0.02 mag bin, and with $V \sim$ 17 mag).   We extrapolate the
fit to fainter magnitudes and subtract it from the data.
This reveals a clear ``excess'' of stars on the giant branch
totaling $a \approx 22000$ stars.
The observed ratio of AGB-bump to RR~Lyrae
stars in the 9M~CMD is $a/v \sim 1.5$, which is nearly eight times
the $M = 0.75 M_{\odot}$ model prediction.  

Adopting the ratio of AGB-bump to HB stars
$a/(b+v+r) \approx 0.05$ from the models and assuming
$b << (v+r)$, the ratio
$a/v \sim 1.5$ yields $r/v \approx 30$, or 
$(b-r)/(b+v+r) \approx -0.97$.   We compare this
HB morphology index to those of ancient LMC clusters 
in the next section.
Scaled by the
number of RR~Lyrae variables, this HB morphology predicts
$r \approx 455000$ red HB clump giants, which is probably uncertain by a
factor of 2.  In any case, $r$ is a very large number.
Finally, we estimate the 
total number of red HB clump stars in the 9M~CMD to be
$r \sim 900000$ (the measurement procedure was similar to that 
described for the AGB-bump).
By inspection of Figure~6, we conclude that $\sim$455000 red
HB clump giants associated with the old and metal-poor 
field population could only lie under the
main peak of the observed red HB clump, 
which therefore likely indicates
their mean brightness.

\subsection{Summary and Discussion of Old Populations}

The field of the LMC bar is composed, 
in part, of an old population 
that produces RR Lyrae variable stars.
This population appears to have
a metallicity of $[Fe/H] = -1.6$, as evidenced by 
spectroscopy of RRab (Alcock et al.~1996), the distribution of RRab in
the Bailey diagram, and the color of the giant branch in the 9M~CMD.   
In some respects, the Galactic globular cluster M3 is a satisfactory
template for this old and metal-poor population.  However, M3
does not have enough bright red HB clump stars,
and its AGB-bump is bluer 
than the AGB-bump in the 9M~CMD.
It is likely that some of the bright
red HB clump stars in the LMC bar represent an intermediate-age
population.  We suggest that the 1.5 Gyr old,
$[Fe/H] = -0.7$ cluster NGC~411 is a suitable template
for this population.   
The AGB-bump for an NGC~411-like population\footnote{An AGB-bump in NGC~411 
has not been observed, which is probably consistent with the small number of
bright giant branch stars in the cluster.}
would be redder than the AGB-bump in M3, but much brighter
than the AGB-bump in the 9M~CMD.
Thus, while a
simple model for old populations in the 9M~CMD might consist
of two components similar to the clusters M3 and NGC~411, we
find some discrepancies in detail.

The highly-evolved AGB in the 9M~CMD shows three prominent features:
(1) a concentration of stars near the tip-RGB 
consistent with an old and
metal-poor population like NGC~1898, (2) an extended faint
sequence which lacks carbon stars, and (3) an extended
bright sequence which runs parallel in the 9M~CMD to the more prominent 
faint sequence.  
The bright sequence is consistent with an intermediate-age
population like NGC~1783 (or NGC~411).
The population associated with this bright AGB sequence 
may account for the majority of LMC carbon stars.
We suggest that the multiple AGB sequences in the 9M~CMD 
represent discrete old populations in the LMC bar.  The concentration
of stars near the tip-RGB and those on
the faint sequence may be old and metal-poor,
while those on the bright AGB-sequence 
may be significantly younger, and relatively
more metal-rich.  However,
inferring the nature of stellar populations from the AGB is difficult 
because the predictions 
of stellar evolution theory are
uncertain for these highly evolved stars,
and few AGB stars are observed in
suitable template clusters.

There are several indications that the old and metal-poor field population
in the 9M~CMD has a red HB morphology: the AGB-bump is quite red,
the field RRab lie close to the red edge of the instability strip,
and the number ratio of AGB-bump to RR Lyrae stars 
is quite large.  For an HB morphology of
$(b-r)/(b+v+r) \approx -0.97$, the ratio $a/v \approx 1.5$ associates
$\sim$50\% of the red HB clump stars in the 9M~CMD
with an old and metal-poor population (the same population responsible
for the RR Lyrae).  

The red HB clump giants in the 
9M~CMD are distinctly brighter than the red HB stars in M3
relative to the RR~Lyraes,
which may suggest the relative youth of the former. 
If we adopt a canonical brightness difference between RRab and a red HB
of $-0.1$ mag, as found for globular clusters (Fusi Pecci et al.~1996),
then the red HB clump in the 9M~CMD is about $-0.2$ mag
brighter than a typical, ancient red HB.  
If ancient clusters are $\sim$13 Gyrs old, 
theoretical calibrations of the age-dependent red HB clump
luminosity predict that the
old and metal-poor field population in the bar 
is $\sim$5 Gyrs old (Alves \& Sarajedini 1999, Sarajedini et al.~1995).
A systematic uncertainty in this
estimate may arise from the possible coupling of
age and metallicity effects on the 
theoretical red HB clump luminosity-age calibration, 
the nature of which is poorly constrained
by observation.

In addition to the
brightness of the red HB clump, the inferred
HB morphology index of $(b-r)/(b+v+r) \simlt -0.97$
may also indicate the relative youth of the metal-poor field
population.
For $[Fe/H] = -1.6$ dex, 
the Lee et al.~(1994) models predict that the
field population is at least $\sim$2 Gyr younger than the oldest LMC
clusters (Olsen et al.~1998).
It is worth noting the magnitude of the
age difference inferred from the Lee et al.~(1994)
models depends sensitively on a variety of theoretical
assumptions (Catelan \& de Freitas Pacheco 1993).  
Nevertheless,
the weight of evidence supports a scenario whereby the
majority of old and metal-poor field stars 
in the bar formed {\it after}
the oldest LMC clusters.

The brightness difference
between the red HB clump and AGB-bump in the 9M~CMD,
$\Delta V^{Bump}_{Clump} \approx -0.8$ mag, does not agree
well with observations of old and metal-poor clusters
(e.g.~Ferraro et al.~1999), or theoretical predictions. 
If the old and metal-poor field population
is as young as $\sim$5 Gyr, the disagreement with
theory is even worse (Alves \& Sarajedini 1999).
However, we find that the brightness difference 
between the AGB-bump and RRab
in the 9M~CMD
is in good agreement with 
observations of globular clusters and with theory.
Therefore, while the brightnesses of the RRab
and AGB-bump conform to expectations,
the red HB clump appears too bright.

To review, we find that the brightness difference
between the AGB-bump and RR~Lyraes is consistent with the luminosity
predictions of stellar evolution theory for an old and metal-poor
population.
The observed ratio of 
AGB-bump stars to RR~Lyraes and the lifetime
predictions of stellar evolution theory lead us to associate a
significant fraction of the observed red HB clump giants with
this old and metal-poor population, which in turn, allows us to
infer their mean brightness.  
However, these red HB clump giants appear to be brighter relative 
to the AGB-bump and RR~Lyraes
than predicted by stellar evolution theory.  We have argued 
that the old and metal-poor field population is probably younger
on average than the truly ancient LMC clusters.  However,
this only worsens
the disagreement with theory.

Have we incorrectly
associated a large fraction of the bright red HB clump giants in the 9M~CMD
with an old and metal-poor population?  
One possibility is that an RGB-bump of an intermediate-age
population may also populate the region of the
giant branch near the AGB-bump in the 9M~CMD.  
A young RGB-bump like this may have been observed in
NGC~411, and is predicted by theory
(Alves \& Sarajedini 1999).  If we have overestimated the number 
of AGB-bump stars, we would expect
a smaller number of old, faint red HB clump giants in
the 9M~CMD.  (These old, faint red HB clump giants may be
overwhelmed by a much larger number of young
red HB clump giants.)
However, this scenario probably requires
far too many $\sim$1.5 Gyr old stars 
populating a very bright RGB-bump (Hardy et al.~1984).
In addition, we do not observe a bright
AGB-bump, which would be expected if a $\sim$1.5 Gyr 
population dominated the 9M~CMD giant branch.

Are the HB models missing 
a parameter other than age that would produce bright
red HB clumps 
in old and metal-poor populations?
In this case, the hypothesized parameter
would not appear to affect the luminosities
of the AGB-bump or RR~Lyraes.  
A second possibility is that the luminosity
of the red HB clump has a very strong dependence on age in metal-poor
populations (i.e., a strong coupling of the age and metallicity dependencies 
of the red HB clump luminosity).  
This trend is seen in the HB models of Sarajedini et al.~(1995), but
the effect is probably too small 
to account for the $\sim$0.2 mag found here.
In either of these scenarios,
the old and metal-poor field
population in the LMC bar may still be younger than the 
oldest clusters, but the
currently available HB models would not yield 
accurate estimates of the age difference.   Finally, we note
that the inferred brightness of these old and metal-poor red HB clump 
giants follows directly from the
lifetime and luminosity predictions of stellar evolution theory 
for the AGB-bump.
The possibility that these predictions 
are flawed seems
unlikely given the good agreement between the observed 
and theoretically-predicted values of $\Delta V^{Bump}_{RRab}$.
New theoretical models 
and further observational detections 
of AGB-bumps in clusters 
might help clarify this situation.

\section{The Old LMC Disk}

We conclude this work
with a brief examination 
of the spatial density profile of RRab.  
In Figure~12, we plot the
logarithm of the number of 
RRab per square degree 
as a function of true LMC radius.  We assume the LMC
disk is inclined ($i = 35$ degrees), with a line-of-nodes
running North-South, and a center near the optical
center (Westerlund 1997).  MACHO data for 16 fields
are shown as filled circles (no corrections for completeness).
We plot data for six additional fields 
from Kinman et al.~(1991) as open circles
(error bars are $\sigma_{S} = 2S^{1/2}$).  
A fit to all of the
data (dotted line)
and a fit excluding the Kinman et al.~(1991) data (solid line)
are also shown. 
We report a radial scale length of $\Lambda = 1.6 \pm 0.1$ kpc
(statistical error only), which is the same
as derived from surface brightness data (Bothun \& Thompson 1989).
For the purposes of this work, we need only illustrate the good fit
of the exponential profile. 
Fits with analytic King (1962) models (not shown) are 
worse than the exponentials.
The King models tend
toward vanishingly small core radii, and are
poor representations of the Kinman et al.~(1991) data points.
Future work might include a more thorough exploration
of the model parameter space (i.e., accounting for uncertainties 
in the LMC center, 
inclination, and line-of-nodes), and 
corrections for RR~Lyrae completeness as estimated from
artificial star tests (Alcock et al.~2000).
In any case, the 
data shown here 
strongly suggest that the majority of the old and metal-poor LMC field
stars lie in a disk, and not in a spheroid.

A lower limit for the age
of the LMC disk is set by the age of the
youngest RR~Lyrae, which may be $\sim$ 9 Gyr following
the discussion of Olszweski et al.~(1996).  An upper limit
is set by the old and metal-poor LMC field population 
forming after the
oldest LMC clusters (as we have argued here), 
and the absolute ages of those
clusters.  We suggest a plausible range of
$9 \simlt \tau_{disk} \simlt 12$ Gyr for the formation
epoch of the LMC disk.
The Milky Way disk is estimated to be $\sim$9 Gyr old
(Leggett et al.~1998, Salaris \& Weiss 1998) from white dwarf
cooling times and the ages of disk clusters.  Therefore, we
conclude that the LMC disk is quite old, and may have
formed contemporaneously with the 
Milky Way disk.

\section{Conclusion}

We have presented a 9 million star color-magnitude diagram 
of the LMC bar.
By assembling a variety of theoretical 
results, carefully selected
samples of MACHO-discovered variable
stars, and extant observational data for clusters (Galactic, LMC, 
\& SMC), we have investigated the stellar populations of the LMC bar.
After an examination of the young LMC stellar populations, we turned
our attention to the less well-understood 
old stellar populations.   
Regarding the old and metal-poor
LMC field population, we found
a typical metallicity of $[Fe/H] = -1.6$, and argued
that this
population likely formed a few Gyr after the oldest LMC
clusters.  We showed that this population lies in a disk.
Our results
are consistent with isochrone-dependent parametric studies of
stellar populations in the LMC disk 
(Geha et al.~1998), which do not rule out a
substantial $\sim$10 Gyr old and metal-poor population.
We are also consistent with
recent suggestions that the LMC bar harbors a relatively
larger component of old stars than the outer regions of the LMC disk
(Holtzman et al.~1999).

As emphasized by others (Butcher 1977, Hardy et al.~1984), 
the LMC disk appears to have a different
star formation history than the solar neighborhood Milky Way disk.
The key difference hinges on a major ``event'' of 
star formation which apparently took place 
a few Gyr ago in the LMC.   
The details of this event and the prior evolutionary
history of the LMC are the subject
of some debate (Bertelli et al.~1992, Olszweski et al.~1996,
Elson et al.~1997, Geha et al.~1998, Holtzman et al.~1999).
We have suggested that the bimodal AGB in the 9M~CMD arises from discrete old
populations, which is consistent with the occurence of a major star formation
event in the LMC.
These discrete populations may be due to either
a distinguishing ``dip'' in the star formation rate prior to
the event, or a ``burst'' coincident with the event.
In addition, we showed that the red HB clump, giant branch, and AGB 
in the 9M~CMD are
consistent with the so-called burst population having a
metallicity and age similar to the clusters NGC~411 and NGC~1783,
which is somewhat younger and more metal-poor than others
have suggested.

Although our results are consistent with the occurence of a
star formation event a few Gyrs ago in the LMC, we do not
constrain the intensity of the event.
Consider the following
simple calculation. 
If $\sim$50\% of the red HB clump giants are old and metal-poor,
we assume that the remaining
$\sim$50\% are born in the event/burst.
Adopting equal red HB clump lifetimes
(Vassiliadis \& Wood 1993), a Salpeter (1955) initial mass function,
and representative 
initial masses of 0.8 and 1.6 $M_{\odot}$ for the old and
intermediate-age populations, respectively,
the roughly equal numbers of red HB clump giants
from the two populations
imply relative star formation rates of $SFR_{1.6}/SFR_{0.8}
\propto (0.8/1.6)^{-2.35} \sim 5$.  
If the number of old and metal-poor clump giants 
is uncertain by a factor of $\sim$2, and the total number
of red HB clump giants is fixed, 
the ratio of star formation rates may range from
1 to 10.

Finally, we have shown that the old and metal-poor field population in the
LMC bar lies in a disk.   This contradicts suggestions
that the LMC disk formed only a few Gyr ago, during the star formation 
event (Hardy et al.~1984, Elson et al.~1997).
Instead, we find that the 
LMC disk is at probably as old as the
Milky Way disk, $ \tau_{disk} \sim 9$ Gyr.
The Milky Way and the LMC have absolute visual magnitudes
of $M_{V} \approx -20.4$ and $-18.4$ mag, and masses
of $M \approx 6\times10^{10}$ and $3\times10^{9}$ $M_{\odot}$,
respectively (Bahcall \& Soneira 1980, 
de~Vaucouleurs 1954,
Bothun \& Thompson 1988,
Kim et al.~1998).
Thus, we conclude that very different mass galactic disks may form 
at similar and quite early cosmological epochs, 
a new clue which may help constrain
the physical processes governing their formation.
It is interesting to note that the kinematics
of the ancient LMC clusters do not reveal
a halo, but rather a disk (Freeman, Illingworth \& Oemler 1983).
The existence of a bonafide
spheroidal or halo stellar population in the LMC
is not ruled out by our analyses.   
The metal-poor RR~Lyraes in the tail 
of the metallicity distribution 
are likely good candidates for this LMC
population.
Identifying this
population will be important for
probing the pre-disk/collapse era of galaxy evolution
in the LMC (Eggen, Lynden-Bell \& Sandage 1962).

\clearpage

\section{Acknowledgements}

This paper is excerpted from the Ph.D. dissertation
by David~Alves for the Department of Physics, University
of California, Davis.
David Alves thanks his graduate advisors, Dr.~Kem Cook
and Prof.~Robert Becker.
The MACHO collaboration 
thanks the skilled support by the technical staff at MSSSO.
Work at LLNL supported by DOE contract W7405-ENG-48.  Work at CfPA 
supported by NSF AST-8809616 and AST-9120005.  Work at MSSSO supported
by the Australian Dept.~of Industry, Technology and Regional Development.
WJS thanks PPARC Advanced Fellowship, KG thanks DOE OJI, Sloan, and
Cottrell awards, CWS thanks Sloan and Seaver Foundations.

\clearpage
\singlespace
\normalsize
\footnotesize

\begin{deluxetable}{ll|ll}
\tablewidth12cm
\tablecaption{Fiducial Marks}
\tablehead{ 
\colhead{$V$} &
\colhead{$(V-R)$} & 
\colhead{$V$} &
\colhead{$(V-R)$} 
}
\startdata
Blue Supergiants\tablenotemark{a} & \ & Red Supergiants\tablenotemark{a} \nl
\ \  14.75 & 0.09 & \ \ 14.75 & 0.76 \nl
\ \  15.25 & 0.18 & \ \ 15.25 & 0.71 \nl
\ \  15.75 & 0.30 & \ \  15.75 & 0.66 \nl
\ \  16.25 & 0.39 & \ \ 16.25 & 0.64 \nl
\ & \ & \ & \ \nl
Field RRab\tablenotemark{b} & \ & Red HB Clump\tablenotemark{b} & \ \nl
\ \  19.45 & 0.31 & \ \  19.17 & 0.51  \nl
\ & \ & \ & \ \nl
Giant Branch\tablenotemark{c}   & \  & AGB-Bump\tablenotemark{b} & \  \nl
\ \  16.75 & 0.825 & \ \ 18.38 & 0.57 \nl
\ \  17.25 & 0.713 & \ \nl
\ \  17.75 & 0.639 & Tip-RGB & \ \nl
\ \  18.25 & 0.576 & \ \ 16.26 & 0.80 \nl
\ \  19.75 & 0.505 & \ \ 16.76 & 0.95 \nl
\enddata
\tablenotetext{a}{\footnotesize
Median color for stars with $V \pm 0.25$ mag. 
Estimated uncertainty $(V-R)$ is 0.01 mag.  See text for data selection. }
\tablenotetext{b}{\footnotesize
Estimated uncertainty $(V-R)$ \& $V$ is 0.01 mag.}
\tablenotetext{c}{\footnotesize
Twenty two field average (of median color per field) for stars
with $V \pm 0.25$ mag.
Standard deviation $\sigma_{V-R} \approx 0.04$ mag for
each mark gives width of giant branch. See text for data selection.}
\end{deluxetable}

\clearpage
\begin{landscape}

\begin{deluxetable}{lllllll}
\tablewidth15cm
\footnotesize
\tablecaption{NGC~1898 RRab \tablenotemark{a} }
\tablehead{ 
\colhead{MACHO Id.} &
\colhead{RA} &
\colhead{Dec.} &
\colhead{Period} & 
\colhead{$V$} &
\colhead{$V-R$} &
\colhead{$A_V$} 
}
\startdata
78.5857.2941 & 05:16:46.7 & $-69$:39:07 & 0.475227  & 19.336  & 0.287  & 1.16   \nl
78.5857.2419 & 05:16:46.6 & $-69$:39:24 & 0.492045  & 19.274  & 0.300  & 1.25   \nl
78.5857.2749 & 05:16:29.0 & $-69$:39:27 & 0.532596  & 19.718  & 0.334  & 1.21   \nl
78.5857.2984 & 05:16:38.5 & $-69$:40:11 & 0.554789  & 19.215  & 0.243  & 1.00   \nl
78.5857.2200 & 05:16:33.2 & $-69$:38:23 & 0.647280  & 19.212  & 0.414  & 0.55   \nl
   & & & & & &   \nl
{\it Mean }   & & &  0.540  &  19.351  & 0.315  & 1.04   \nl
{\it S.D.o.M. }  & & & 0.029  & \ 0.095 &  0.029 & 0.13  \nl
\enddata
\tablenotetext{a}{\footnotesize RA \& Dec. are J2000.
Period is given in days; $V$, $V-R$,
$A_V$, and $A_R$ are in mag.}
\end{deluxetable}
\end{landscape}

\clearpage
\begin{landscape}

\begin{deluxetable}{llllllll}
\tablewidth15cm
\footnotesize
\tablecaption{NGC~1835 RRab\tablenotemark{a} }
\tablehead{ 
\colhead{MACHO Id.} &
\colhead{RA} &
\colhead{Dec.} &
\colhead{Period} & 
\colhead{$V$} &
\colhead{$V-R$} &
\colhead{$A_V$} &
\colhead{Id.\tablenotemark{b}} 
}
\startdata
1.4046.765  & 05:05:14.7 & $-69$:24:12 & 0.517913  & 19.479 & 0.214 & 1.32 &  21 \nl
1.3925.1806 & 05:04:46.2 & $-69$:22:27 & 0.525116  & 19.149 & 0.157 & 0.96 &   \nl
1.4046.809  & 05:05:13.1 & $-69$:23:57 & 0.541014  & 19.381 & 0.308 & 1.06 &  18 \nl
1.4046.783  & 05:05:06.1 & $-69$:25:01 & 0.546220  & 19.197 & 0.180 & 1.12 &  14 \nl
1.4046.937  & 05:05:02.2 & $-69$:24:39 & 0.555905  & 19.626 & 0.242 & 1.18 &  13 \nl
1.4046.858  & 05:05:12.8 & $-69$:24:40 & 0.603043  & 19.496 & 0.238 & 1.17 &  20 \nl
1.3925.1752 & 05:04:58.6 & $-69$:23:34 & 0.635710  & 19.417 & 0.306 & 1.02 &  6 \nl
1.4046.715  & 05:05:02.1 & $-69$:24:36 & 0.664359  & 19.397 & 0.174 & 0.81 &  \nl
 & & &  & & & &  \nl
{\it Mean }  &  & & 0.581  &  19.428 & 0.237 & 1.10 & \nl
{\it S.D.o.M. }  & & &  0.020  & \ 0.049 &  0.020 & 0.06  & \nl
\enddata
\tablenotetext{a}{\footnotesize RA \& Dec. are J2000.
Period is given in days; $V$, $V-R$,
$A_V$, and $A_R$ are in mag.}
\tablenotetext{b}{\footnotesize Identifcations from Walker (1993) 
and Graham \& Ruiz (1974).}
\end{deluxetable}

\end{landscape}

\clearpage
\begin{landscape}

\begin{deluxetable}{lllllllll}
\footnotesize
\tablewidth15cm
\tablecaption{LMC Cluster AGB Variables}
\tablehead{
\colhead {MACHO Id.} & 
\colhead {Id.} & \colhead {$<P>$}  &
\colhead {$<R>$} & \colhead {$<V>$} & 
\colhead {$K$} & 
\colhead {$J-K$} & \colhead {$m_{bol}$} & 
\colhead {Sp.T.}
}
\startdata
{\bf \ NGC 1783 } & &  & & & & & & \nl
55.3008.13  & LE-1 & 410  & 15.77 & 17.55 & 10.26 & 1.93 & 13.47 & C    \nl
55.3008.12  & LE-2 & 46.3 & 15.06 & 16.22 & 11.27 & 1.13 & 14.19 & S4/2 \nl
55.3129.16  & LE-3 & 307  & 15.07 & 16.41 & 10.37 & 1.60 & 13.41 & C    \nl
55.3129.12  & LE-4 & 128  & 15.06 & 16.22 & 11.10 & 1.09 & 13.98 & S5/2 \nl
55.3129.14  & LE-5 & 120  & 15.03 & 16.09 & 11.23 & 1.09 & 14.11 & M3   \nl
55.3008.14  & LE-6 & 62.4 & 15.07 & 16.11 & 11.34 & 1.08 & 14.20 & M3   \nl
55.3129.13  & LE-9 & 270  & 15.25 & 16.56 & 10.93 & 1.11 & 13.83 & S5/3 \nl
55.3129.15 & LE-15 & 63   & 14.94 & 15.88 & 11.78 & 1.00 & 14.49 & M1   \nl
{\bf \ NGC 1898}  & & & & & & & & \nl
78.5857.27 & AM-6  & 51.5 & 15.30 & 16.06 & 12.38  & 0.94 & 15.03  & M?  \nl
78.5857.25 & AM-4   & 46.1 & 15.32 & 16.20 & 11.71: & 1.08 & 14.62: & M? \nl
78.5857.38 & AM-1  & 35.8 & 15.50 & 16.24 & 12.58  & 0.98 & 15.32  & M?  \nl
\enddata
\end{deluxetable}
\end{landscape}

\doublespace
\small
\normalsize

\clearpage

\begin{figure}
\psfig{file=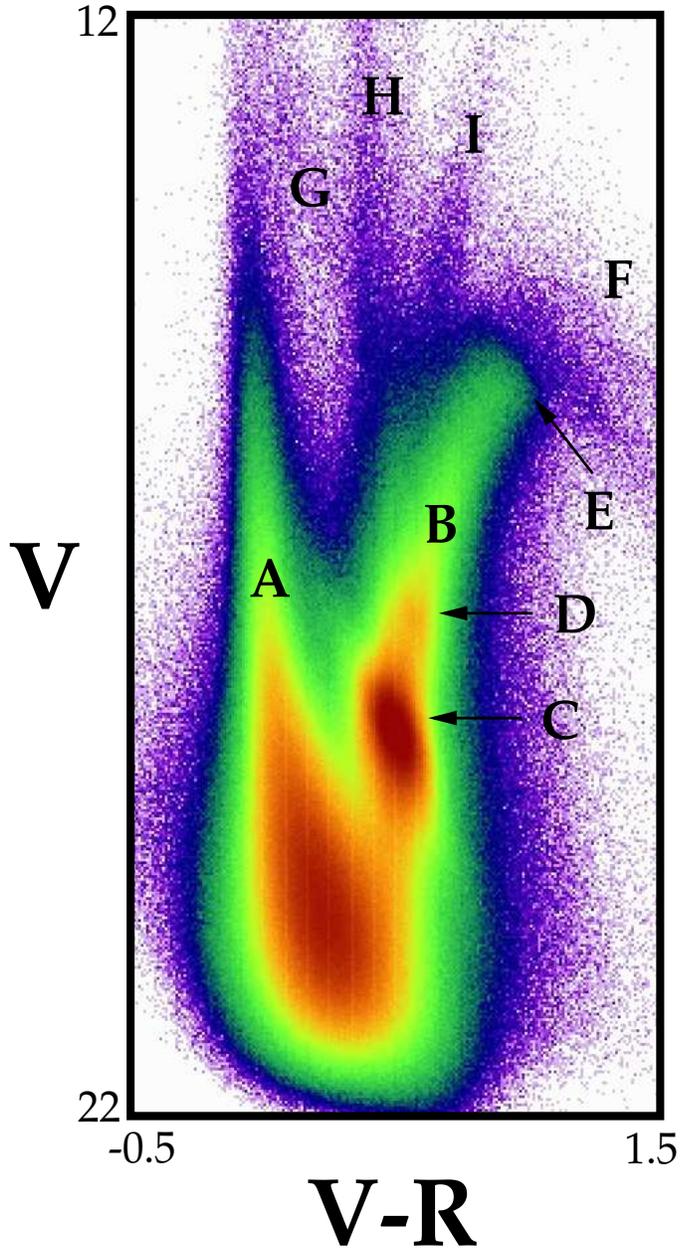,height=6.5in,width=3.5in}
\caption{The 9 million star color magnitude diagram. 
This is a log-scaled and color-coded
Hess diagram where intensity represents the number of stars.  Bin size
is 0.02 mag in $V$ and 0.01 mag in $(V-R)$.  Features discussed in the
text are labeled A thru I. }
\end{figure}

\begin{figure}
\plotone{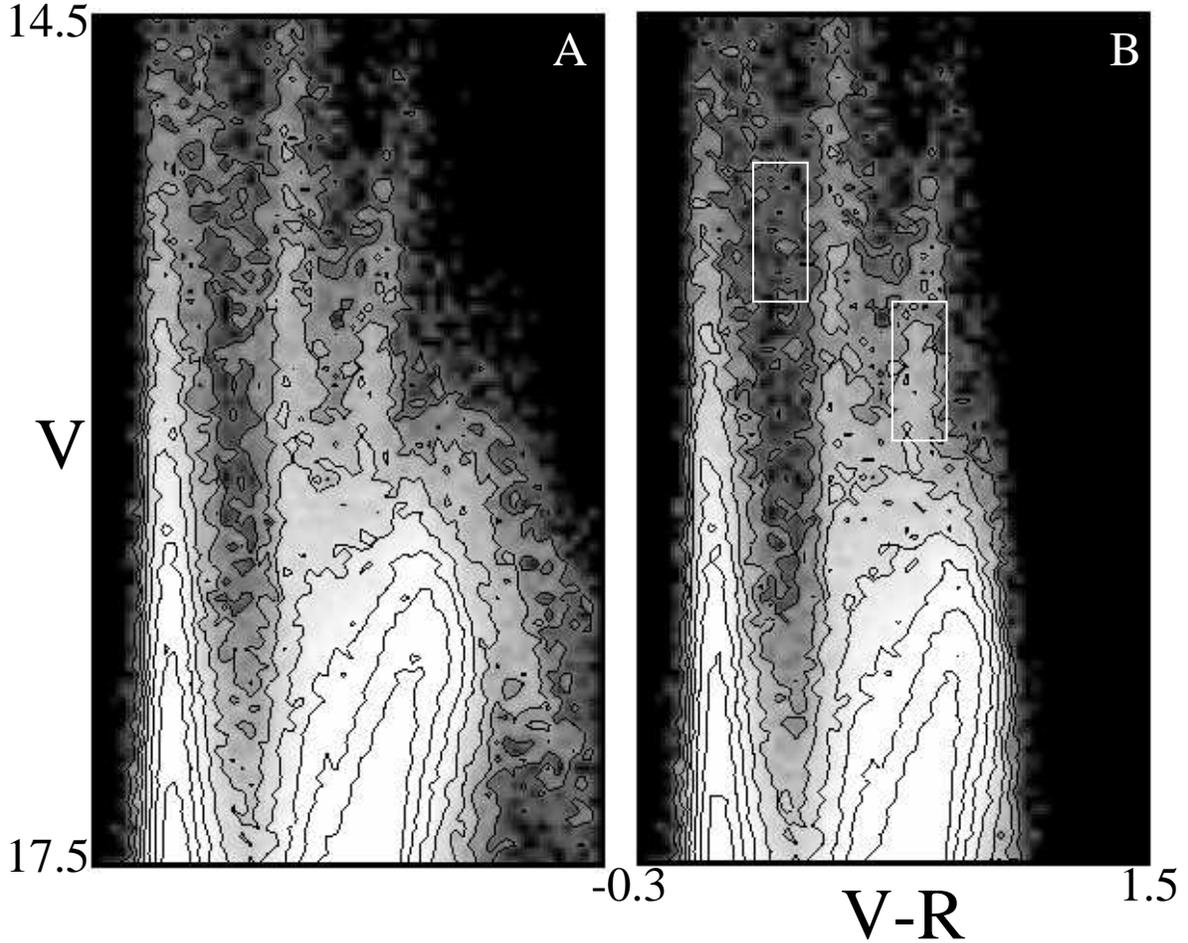}
\caption{Log-scaled Hess diagrams representing 
{\bf (A)} 261805 LMC stars
with $(V-R)$ = $-0.3$ to 1.5 mag, $V$ = 17.5 to 14.5 mag, and 
{\bf (B)} 
the same diagram but with 35169 candidate variable stars removed.  
The variable stars are defined by poor fits to constant brightness
lightcurves.
Intensity and contours 
(1.0 to 2.5 dex in 0.25 steps) indicate the logarithmic number
of stars (no. per $\sim$10 sq.~deg.~and per 0.03 mag square color-mag bin). 
The white square apertures are centered at $(V-R,V)$ = 0.18, 15.25 mag and
$(V-R,V)$ = 0.66, 15.75 mag; they have dimensions $\Delta(V-R,V)$ = 0.2,0.5
mag.}
\end{figure}

\begin{figure}
\plotone{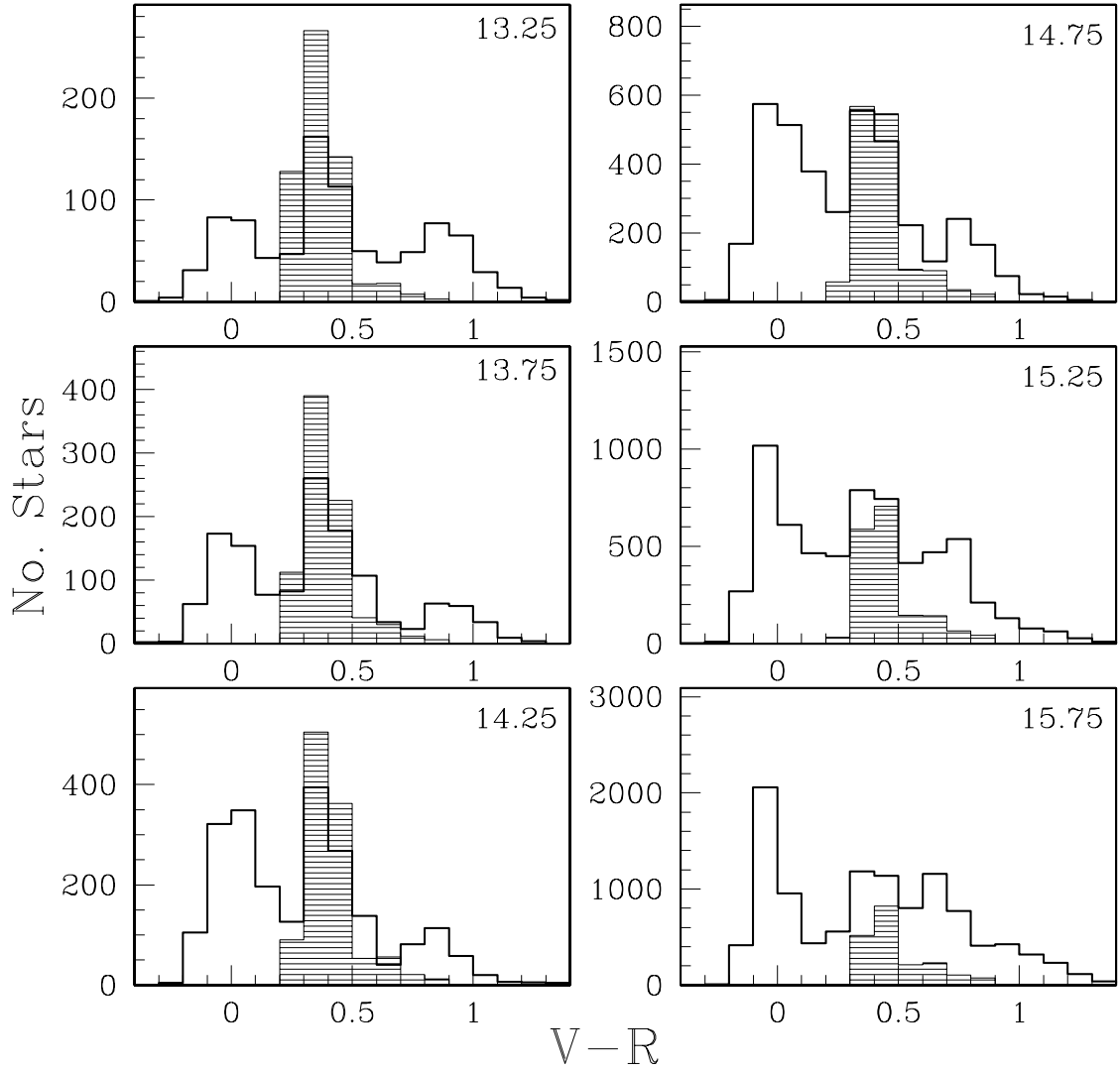}
\caption{Color frequency histograms of the bright stars in the 9M~CMD.
A model prediction for Galactic foreground disk stars is shown as the shaded
histogram.  Each panel shows stars in $\Delta$V = 0.5 mag cuts with the
center $V$ mag labeled in the upper right corner.}
\end{figure}

\begin{figure}
\plotone{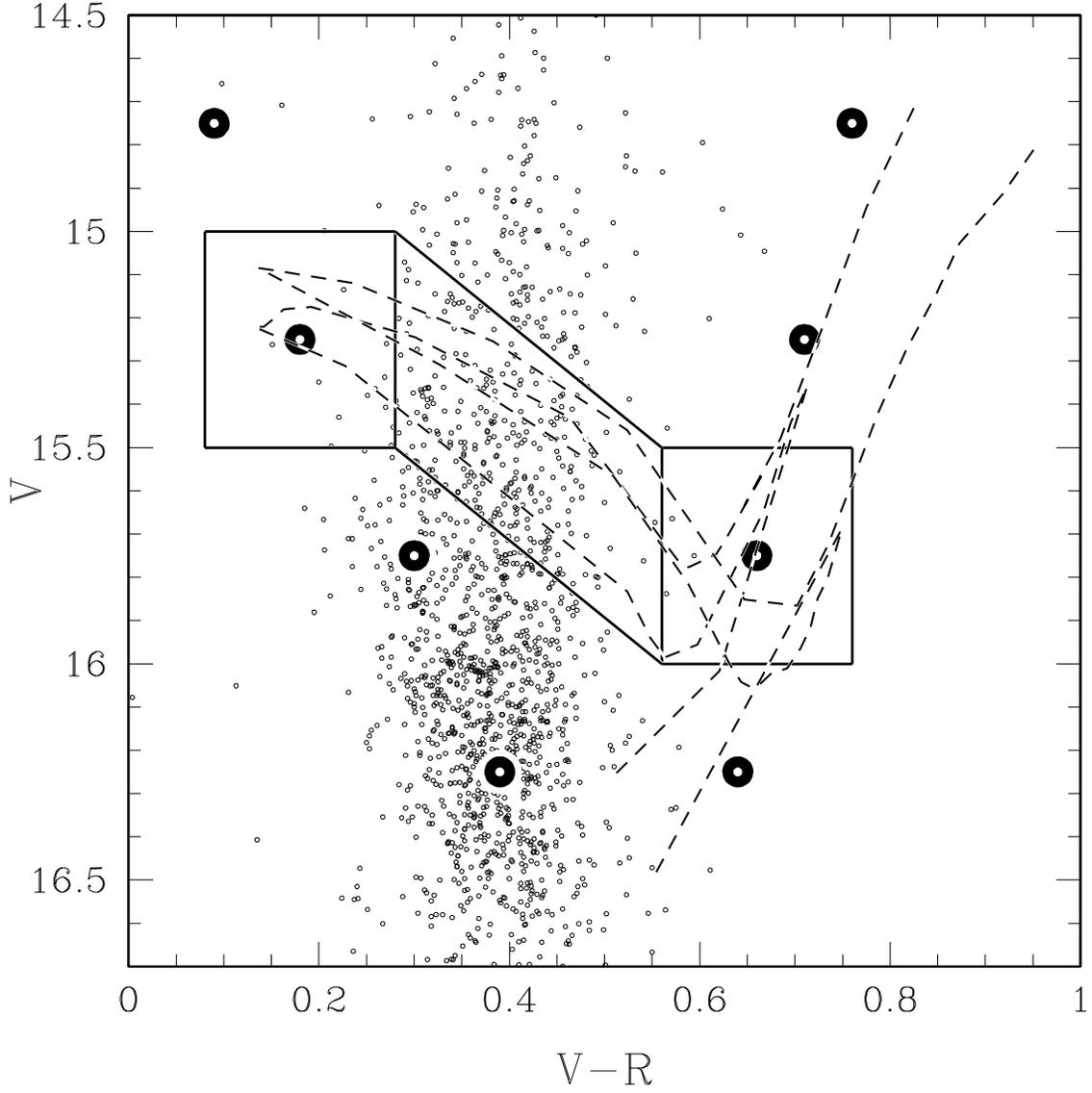}
\caption{CMD showing fiducial marks on 
supergiant sequences (bold circles), apertures
used to count supergiants and Cepheids (solid lines), 
two 5$M_{\odot}$ model sequences (dashed lines; see text), 
and MACHO-discovered Cepheids in the 9M~CMD (small open circles).}
\end{figure}

\begin{figure}
\plotone{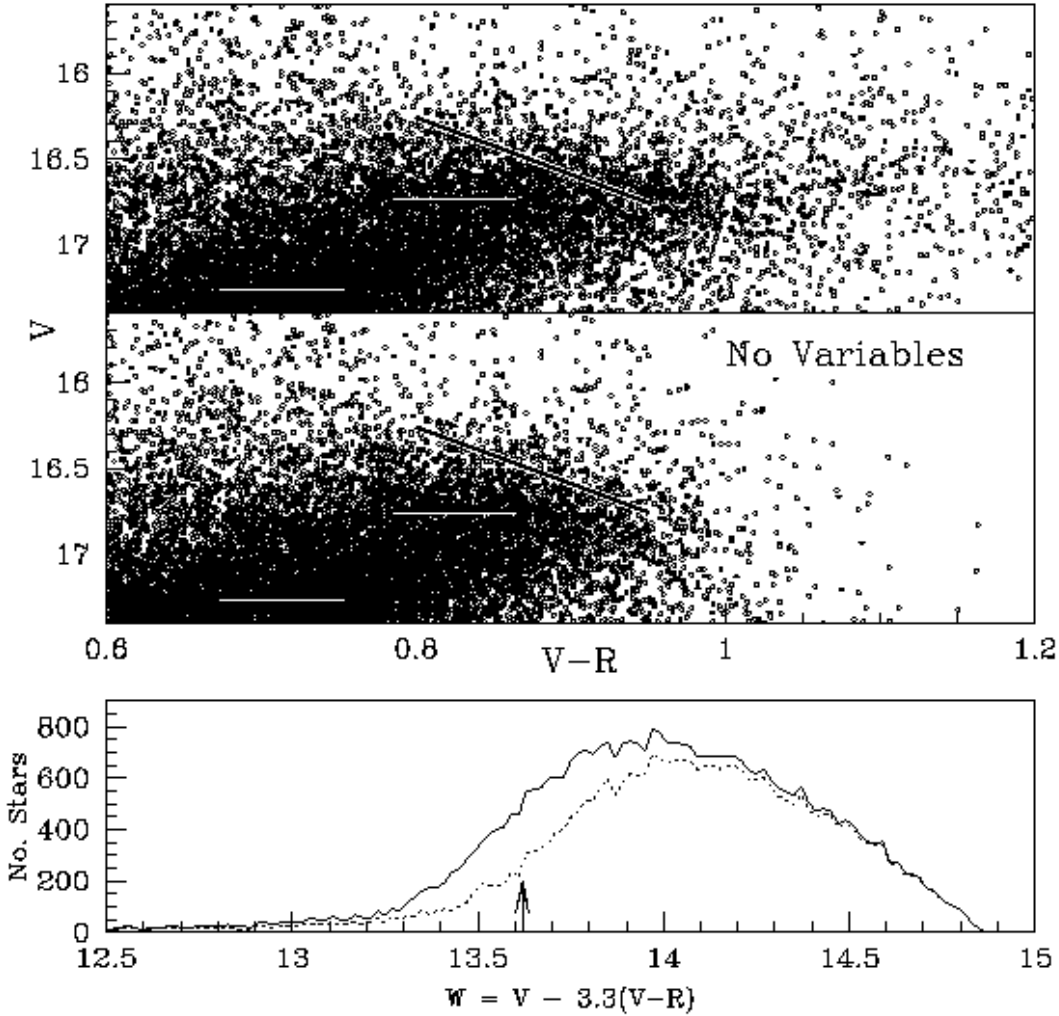}
\vskip-1.7in
\caption{The tip of the RGB in the 9M~CMD. Top panel shows a random
$\sim$10k stars.  Middle panel also shows a 
random $\sim$10k stars, excluding all candidate variable stars. 
The horizontal marks in these two panels are two of the RGB fiducial marks
(see Table~1).  The angled mark corresponds to 
$W_{3.3} = V - 3.3(V-R) = 13.62$ mag
for $0.80 < V-R < 0.95$ mag.  In the 
bottom panel we plot histograms of all stars in this region of the 9M~CMD
(not just the $\sim$10k sample shown in the upper two panels) 
projected along the $W_{3.3}$ vector.
The solid line shows all stars, while the dotted
line shows all non-variable stars.  We find the most significant ``step''
at $W_{3.3} = 13.62$ mag (see text), which is marked with arrow.  
This is the tip-RGB.}
\end{figure}

\begin{figure}
\plotone{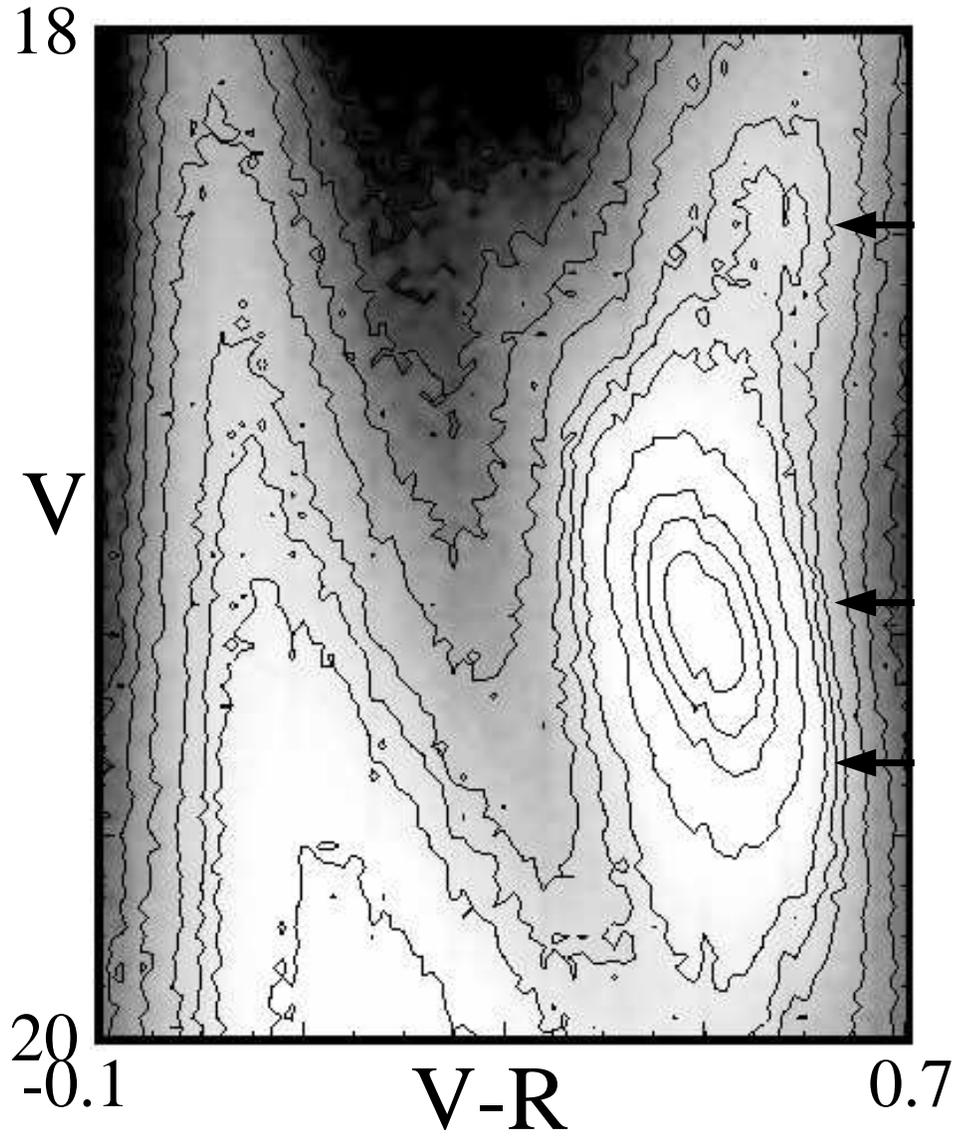}
\caption{Expanded region of the 9M~CMD around the
horizontal branch and AGB-bump.  Logarithmic contours and grey-scale
represent the number density of stars.  Arrows indicate the fiducial
brightnesses of the AGB-bump, red HB clump, and RRab (see text).}
\end{figure}

\begin{figure}
\plotone{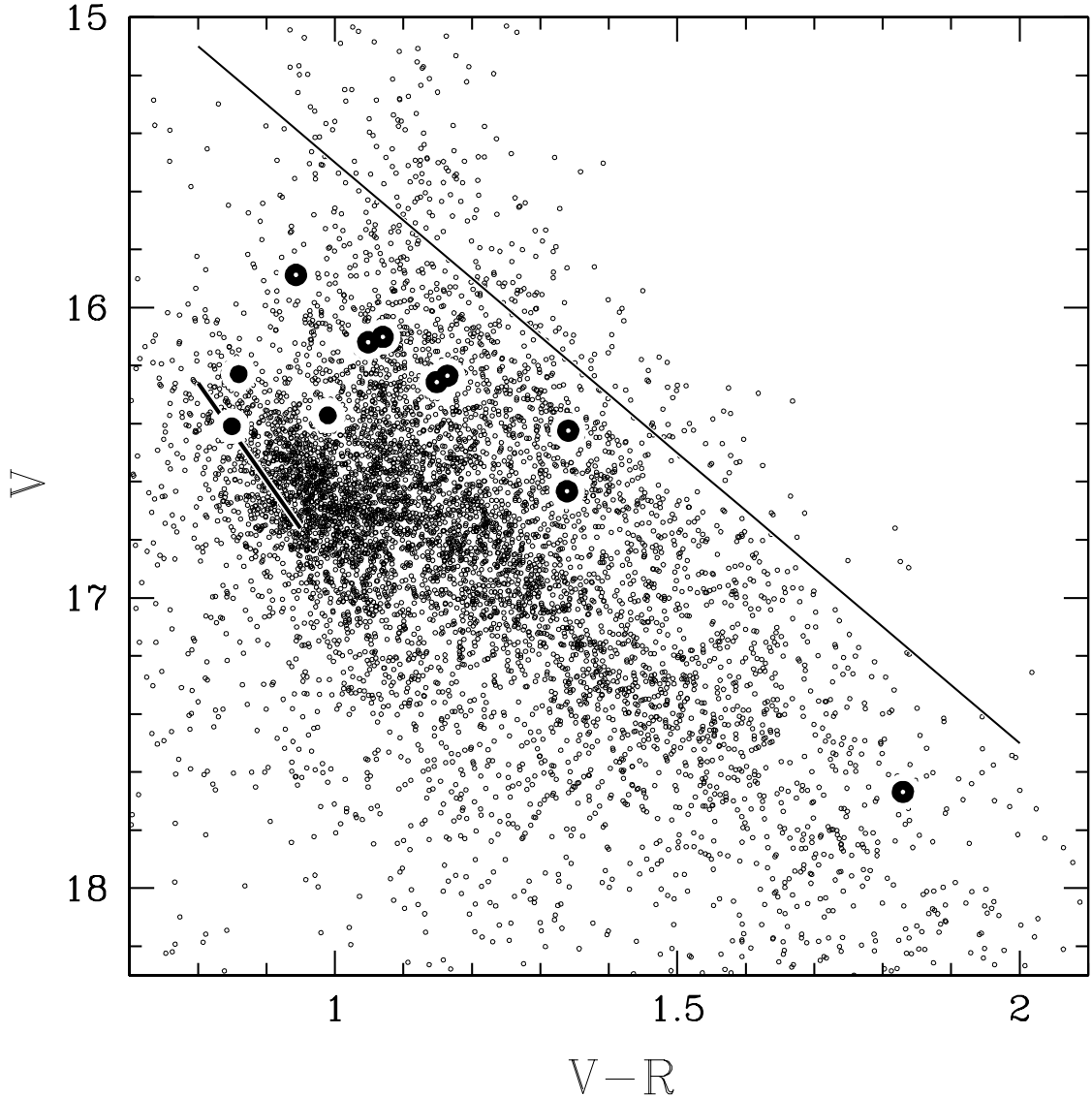}
\caption{This CMD shows a sample of the most statistically significant
variable stars from the 9M~CMD in the region of the AGB.  Several thousand
stars are plotted as small open circles.  We show the tip-RGB as the steeply
angled short mark ($W_{3.3}$) with $0.80 < V-R < 0.95$ mag.  The thin solid
line
corresponds
to $W_{2.0} = V - 2.0(V-R) = 13.5$ mag, which runs nearly parallel to the
extended
sequences of AGB stars.  We use a projection perpendicular to $W_{2.0}$ to 
study the ``thickness'' of the AGB.  The length of this line along the color
axis ($0.8 < V-R < 2.0$ mag) corresponds to the full range of colors 
in Fig.~8.
The bold filled circles are 3 AGB variables from
the LMC cluster NGC~1898.  The bold open circles are AGB variables from the
LMC cluster NGC~1783.}
\end{figure}

\begin{figure}
\plotone{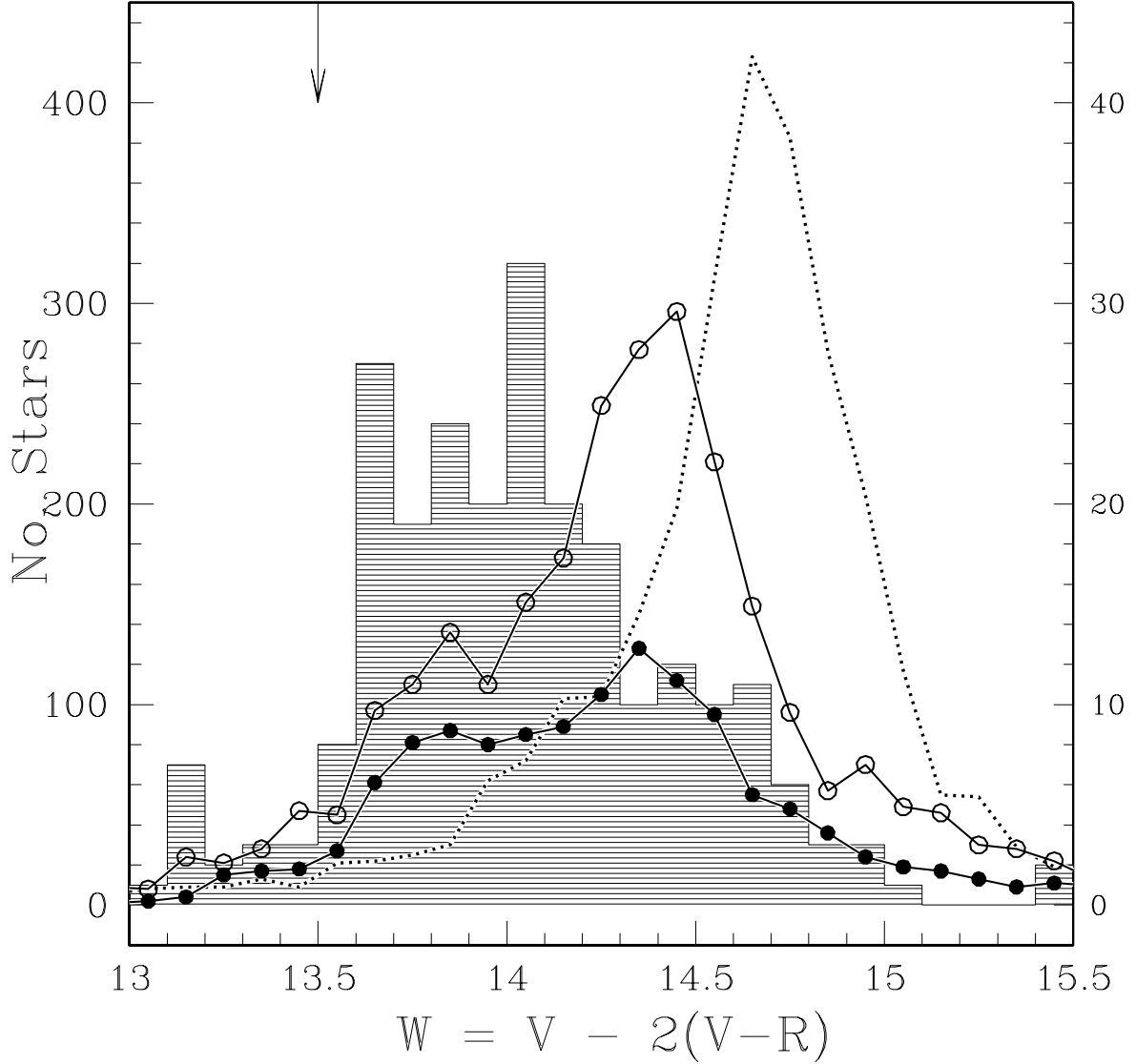}
\caption{Four luminosity functions are shown here, each a projection along
$W_{2.0} = V - 2.0(V-R)$, approximately perpendicular to the sequences
of the AGB in the 9M~CMD.  The same data as
plotted in Fig.~7 are shown 
with three color cuts as follows: the dotted line
is $0.8 < V-R < 1.1$ mag, the solid line with open circles is 
$1.1 < V-R < 1.4$ mag, and the solid line with solid circles 
is $1.4 < V-R < 2.0$ mag.  These $W_{2.0}$ luminosity functions have bin
sizes of 0.1 mag, and the left axis gives the number of stars in each
bin.  The shaded histogram
is the $W_{2.0}$ luminosity function of 266 known carbon stars
(Blanco McCarthy \& Blanco  1985)
identified in the 9M~CMD.
The bin size for the carbon star histogram is also
0.1 mag, and the number of stars is given on the right axis.}
\end{figure}

\begin{figure}
\plotone{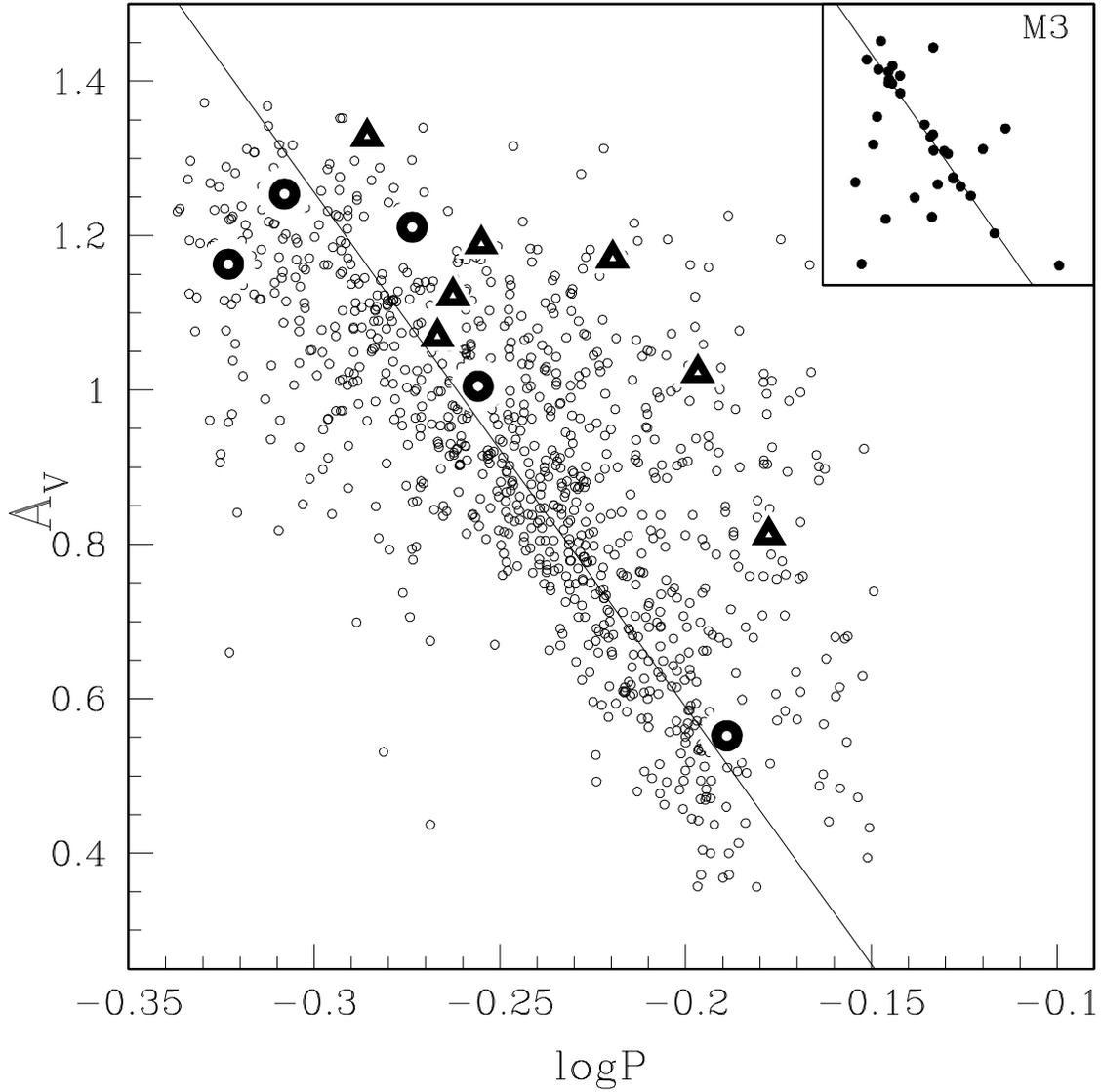}
\caption{Bailey period-amplitude diagram for RRab in the 9M~CMD (small circles)
and RRab in the LMC clusters NGC~1835 (triangles) and NGC~1898
(donuts).  In the inset, we show RRab from the Galactic globular
cluster M3 (filled circles).  The M3 ridge line is indicated
in both panels (see text).}
\end{figure}

\begin{figure}
\plotone{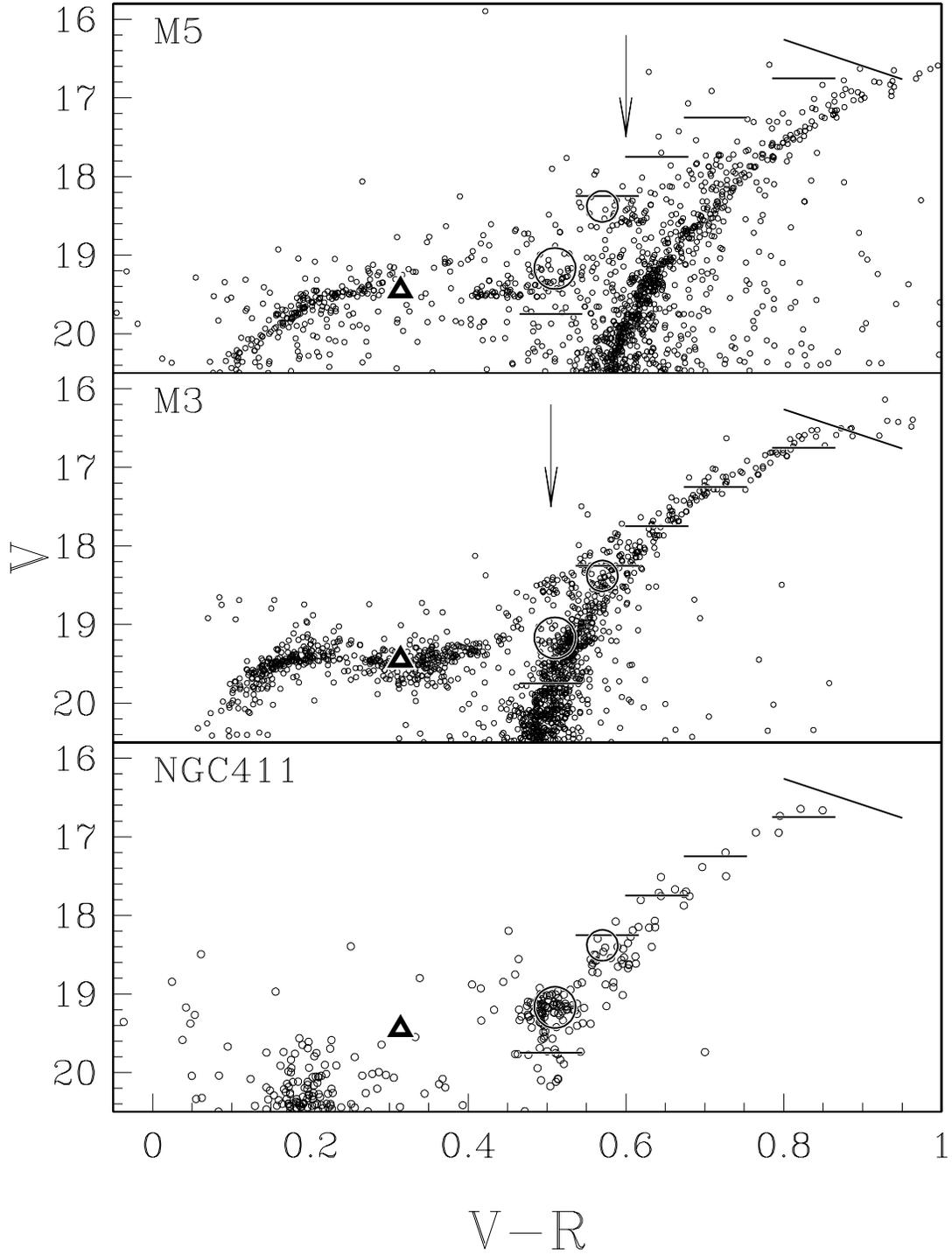}
\caption{In each panel, fiducial marks for the horizontal branch and giant branch 
in the 9M~CMD (see text) are compared with those features in the clusters
with M5, M3 or NGC~411 (small open circles; each panel shows a different cluster).}
\end{figure}

\begin{figure}
\plotone{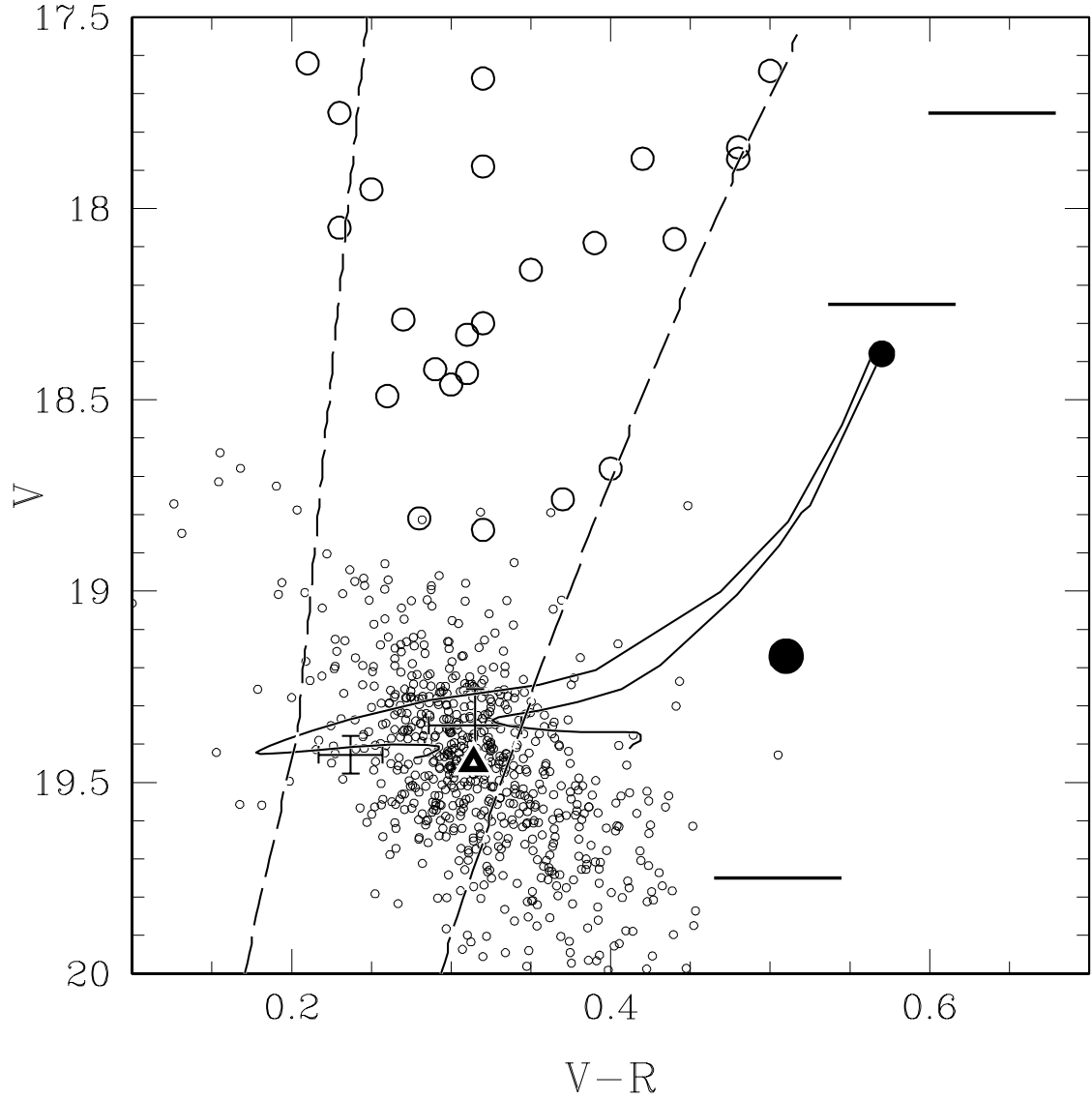}
\caption{Fiducial marks 
for the horizontal branch (RRab and red HB clump),
giant branch (dash marks), and AGB-bump
are compared with a sample of field RRab (small open circles)
and BL~Her variables (large open circles) in the 9M~CMD.  Two HB model
sequences are shown as dotted lines.  
Theoretical edges of the instability strip
are shown as dashed lines.  See text for details.}
\end{figure}

\begin{figure}
\plotone{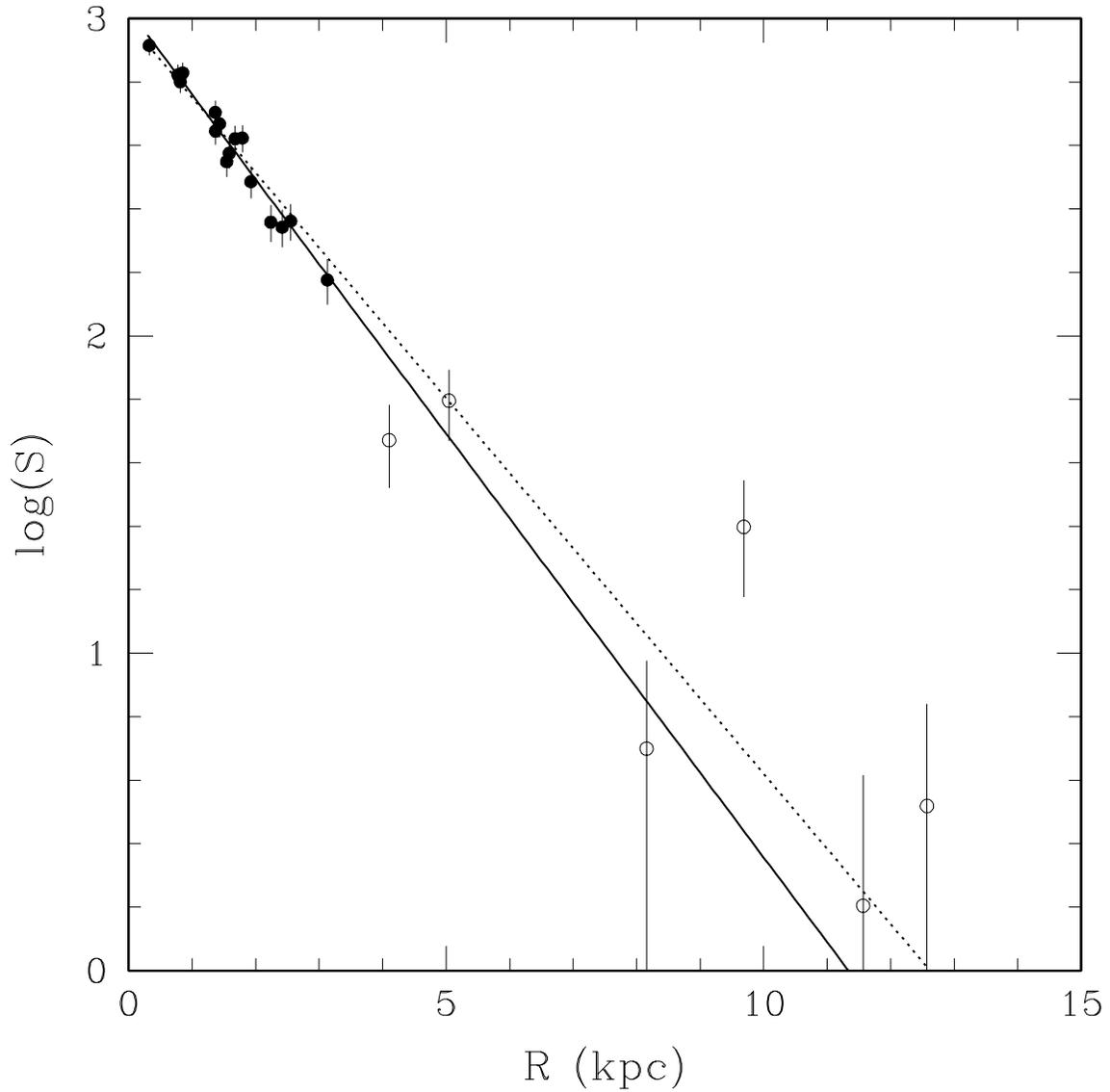}
\caption{Logarithm of the number of RRab per square
degree as a function of true LMC radius (kpc).
MACHO data for 16 fields in and around the LMC bar are plotted as filled
circles.  Data assembled from Kinman et al.~(1991) for six other
LMC fields are plotted as open circles.  Two exponential disk fits
to these data
are plotted as solid and dotted lines (see text).}
\end{figure}

\end{document}